\documentclass{aa}

\usepackage[varg]{txfonts}
\bibpunct{(}{)}{;}{a}{}{,}

\usepackage{amsmath,amssymb,amsfonts}
\usepackage{booktabs}
\usepackage{caption}
\usepackage{graphicx}
\usepackage{subcaption}

\usepackage{tabularx}
\usepackage[english]{babel}
\usepackage[dvipsnames]{xcolor}
\usepackage{aas_macros}

\usepackage{linenoaa}

\usepackage{hyperref}
\hypersetup{
    pdftitle           = {Self-induced confinement of Ultra-High Energy nuclei in cosmic filaments},
    pdfauthor          = {Alessandro Cermanti}
    pdfsubject         = {UHECR},   
    pdfkeywords        = {Astroparticle Physics, Cosmic Rays, Multi-messenger Astrophysics}, 
    pdfpagemode        = {UseOutlines},  
    pdfcenterwindow    = {true},
    pdfdisplaydoctitle = {true},
    pdffitwindow       = {true},
    pdfpagelayout      = {SinglePage},
    pdfstartview       = {FitH},         
    pdfstartpage       = {1},
    pdflang            = {en-US},
    colorlinks         = true,       
    citecolor          = magenta,
    linkcolor          = magenta,        
    filecolor          = blue,   
    urlcolor           = blue,        
    breaklinks         = true,
    bookmarksopen      = true,
    bookmarksnumbered  = true
}

\begin{document}

\title{Self-confinement of ultra-high-energy nuclei in cosmic filaments: implications for the UHECR spectrum and composition}
\titlerunning{Self-confinement of UHE nuclei in cosmic filaments}

\author{Alessandro Cermenati\inst{\ref{inst1},\ref{inst2}}\fnmsep\thanks{alessandro.cermenati@gssi.it} \and Roberto Aloisio\inst{\ref{inst1},\ref{inst2}} \and Carmelo Evoli\inst{\ref{inst1},\ref{inst2}}}
\authorrunning{Cermenati et al.}

\institute{Gran Sasso Science Institute (GSSI), Viale Francesco Crispi 7, 67100 L’Aquila, Italy \label{inst1} \and INFN-Laboratori Nazionali del Gran Sasso (LNGS), via G. Acitelli 22, 67100 Assergi (AQ), Italy \label{inst2}}

\date{\today}

\abstract
{}
{The spectrum and composition of ultra-high-energy cosmic rays (UHECRs) suggest that the population dominating above the ankle releases particles with a suppression at low rigidity, around the EV scale. In phenomenological fits, this feature is usually encoded as an unusually hard source spectrum, which is difficult to reconcile with standard acceleration mechanisms. In self-confinement scenarios, however, such an apparent hardening can arise from transport: escaping UHECRs generate the magnetic turbulence that delays their own release from the magnetized environments surrounding their sources. So far, this mechanism has mainly been explored for its spectral consequences, and its implications for the observed mass composition have not been tested in detail.}
{We extend the self-confinement scenario based on the non-resonant streaming instability to a mixed nuclear composition. The electric current carried by escaping UHECRs amplifies magnetic turbulence in the surrounding intergalactic medium, reduces the diffusion coefficient, and delays particle escape in a rigidity-dependent way. We describe the confined region with an effective leaky-box model including escape, photodisintegration, and secondary production, and we propagate the escaping nuclei to Earth. We then compare the resulting spectrum and composition with Auger measurements and compute the associated cosmogenic neutrino and gamma-ray emission.}
{We find that self-generated turbulence can suppress the escaping flux below rigidities of order $R \sim 1$ EV for source luminosities and magnetic-field coherence lengths compatible with UHECR sources hosted in galaxy clusters and propagating through cosmic filaments. For coherence lengths $\lambda_B \simeq 30$--$50$ Mpc, the model reproduces the main features of the observed UHECR spectrum and composition with a standard acceleration spectrum, $\gamma \simeq 2$, without requiring a dominant primary proton component at the highest energies. During confinement, heavy nuclei efficiently photodisintegrate, producing secondary protons that naturally contribute below the ankle and help account for the observed composition evolution. The same confinement enhances the production of cosmogenic neutrinos and gamma rays: the predicted neutrino flux remains compatible with current limits, while the diffuse gamma-ray background provides a potentially strong constraint on the most extreme confinement configurations.}
{UHECR self-confinement in cosmic filaments therefore offers a physically motivated way to transform standard acceleration spectra into the hard effective spectra inferred from observations. In this picture, the apparent hardness of UHECR sources is not an intrinsic property of the accelerator, but the observable imprint of self-generated magnetic turbulence in the cosmic web.}
{}


\titlerunning{UHECR self confinement near sources}
\authorrunning{A. Cermenati et al.}

\maketitle

\section{Introduction}
\label{introduction}

\nolinenumbers

The energy spectrum and mass composition of ultra-high-energy cosmic rays (UHECRs), as measured by the Pierre Auger Observatory (Auger) at energies \(\gtrsim 6 \times 10^{17}\)~eV, provide the main observational guides to their origin.
The all-particle spectrum~\citep{PierreAuger2020prdspectrum,PierreAuger2021epjcspectrum} exhibits a rich structure, with robust features such as the ankle (\(E_{\rm ankle}\sim 5 \times 10^{18}\,{\rm eV}\)), the instep (\(E_{\rm instep}\sim 1.4 \times 10^{19}\,{\rm eV}\)), and a suppression at high energies (\(E_{\rm sup}\sim 4.7 \times 10^{19}\,{\rm eV}\)).
At the same time, the distribution of the depth of shower maximum, \(X_{\rm max}(E)\), observed by Auger at different energies, provides a proxy for the mass composition of UHECRs and its energy evolution~\citep{PierreAuger2014prdmassa,PierreAuger2014prdmassb,PierreAuger2024icrcfluorecence}.
Joint analyses of the spectrum and composition indicate that the UHECR population across and above the ankle is compatible with a mixed composition of protons and heavier nuclei and a maximum acceleration rigidity of only a few EV.
Assuming that UHE particles escaping from the sources are described by a simple rigidity-dependent power law (\(\propto \mathcal{R}^{-\gamma}\)), these joint analyses generally require very hard spectra \(\gamma \lesssim 1\), and in some cases even inverted ones, \(\gamma < 0\) \citep{PierreAuger2017jcapcombinedfit,PierreAuger2023jcapcombinedfitlowenergy}.

Such hard escape spectra are difficult to reconcile with standard acceleration scenarios, which typically predict \(\gamma \gtrsim 2\), both at non-relativistic and relativistic shocks~\citep{Blandford1987phyrep,Sironi2015ssrv}. Although harder spectra may arise in different settings, such as rapidly spinning neutron stars, stochastic acceleration, or reconnection in relativistic plasmas~\citep{Blasi2000apj,Arons2002apj,Asano2016prd,Winchen2018aph,Sironi2023prl}, the source phenomenology inferred from Auger data strongly suggests that the spectrum released into intergalactic space should be suppressed at low rigidities, and does not directly reflect the acceleration spectrum. This has motivated scenarios in which the UHECR spectrum emitted by the sources is reshaped by transport effects, either during propagation to Earth or in the source environment.

Two general approaches have been discussed in the literature. In the first, the suppression at low rigidity is generated during propagation by diffusion in relatively strong intergalactic magnetic fields, leading to a magnetic-horizon effect~\citep{Aloisio2004apj,Mollerach2013jcap,PierreAuger2024jcapmagnetichorizon}.
In the second, nuclei experience prolonged rigidity-dependent confinement in the source environment, so that interactions with ambient photon fields modify both the escaping spectrum and the mass composition~\citep{Unger2015prd,Muzio2019prd,Muzio2022prd}. Both scenarios can mimic an apparently hard source spectrum, but they require either magnetic fields close to current upper limits or a physical mechanism able to sustain efficient confinement around the sources.

A physically motivated realization of the latter possibility is provided by self-confinement, first proposed in~\citet{Blasi2015prl}. The electric current carried by escaping UHECRs can excite the non-resonant streaming instability (NRSI) in the surrounding medium~\citep{Bell2004mnras,Bell2013mnras}. The resulting self-generated turbulence reduces the diffusion coefficient around the source and can suppress the escaping flux below a characteristic rigidity of the order of 1 EV. In~\citet{Cermenati2026aa}, we developed this scenario in detail and showed that, for plausible source luminosities, \(L \approx 10^{44}\,{\rm erg\,s^{-1}}\), and background magnetic-field strengths, \(B_0 \approx 1\,{\rm nG}\), self-confinement can operate on Mpc scales. It can therefore produce a low-rigidity cutoff in the released spectrum at a rigidity compatible with the phenomenology inferred from Auger data. In that work, we focused on protons and showed that the neutrino flux associated with their enhanced confinement around the sources remains compatible with current IceCube and Auger limits~\citep{IceCube:2025ezc,PierreAuger2024icrcphotons}. That proton-based framework was explicitly presented as a first step toward the mixed-composition case relevant for Auger data.

In this work, we investigate the conditions under which the same mechanism can simultaneously account for the observed UHECR spectrum and mass composition. We extend the self-confinement framework to a mixed nuclear injection, include the relevant interaction channels in the circum-source environment and during extragalactic propagation, and compute the resulting spectrum and composition at Earth. We then compare the model predictions with Auger measurements of the flux and \(X_{\rm max}\) observables.
Our goal is to test whether the hard spectra inferred from phenomenological fits can arise from standard acceleration combined with physically motivated confinement, rather than requiring intrinsically hard acceleration spectra.

We also explore the corresponding multi-messenger signatures as additional probes of the scenario. In particular, for the first time in this context, we compare our predictions for high-energy neutrinos with the flux level suggested by the recently reported extremely high-energy neutrino event detected by KM3NeT~\citep{KM3NET2025prx}, together with the contemporaneous non-observation of similar events in IceCube and at the Pierre Auger Observatory. Finally, we compare our gamma-ray predictions with the extragalactic gamma-ray background measured by Fermi-LAT~\citep{FermiLat2015apj,FermiLat2016prl}.
	
\section{Model}
\label{model}

\subsection{Excitation and growth of the instability}
\label{sec:instability}

We adopt the same physical setup as in~\citet{Cermenati2026aa}, extending it here to a mixed nuclear composition. We consider a steady UHECR source embedded in the intergalactic medium (IGM). As cosmic rays escape from the source, they carry an electric current along the large-scale magnetic field and can excite the non-resonant streaming instability. Since the growth of the instability is controlled by this current, the first step is to relate it to the source luminosity and composition.

For each nuclear species \((A,Z)\), we assume an injection spectrum at the accelerator of the form
\begin{equation}
q_{\rm acc}^{(A,Z)}(E)\;=\;\frac{\eta_{(A,Z)} L}{\Lambda}\,E^{-\gamma}\,
f_{\rm cut}(E/Z,\mathcal R_{\rm max}),
\label{eq::acceleration_spectrum_mixed}
\end{equation}
where \(\eta_{(A,Z)}\) is the fraction of the total luminosity \(L\) injected into nuclei of mass number \(A\) and charge \(Z\), and \(\Lambda\) is the normalization factor. In this paper we assume a spectral behavior of the accelerated particles with \(\gamma \simeq 2\), as in the standard diffusive shock acceleration mechanism.

The cutoff function \( f_{cut} \) depends on the rigidity, \(\mathcal R = E/Z\), and describes the maximum rigidity that the source can provide. We adopt the same shape used in the phenomenological fit of~\citet{PierreAuger2024jcapmagnetichorizon}, fixing \(\mathcal R_{\rm max}=6\)~EV:
\begin{equation}
f_{\rm cut}(\mathcal R,\mathcal R_{\rm max})
\;=\;
{\rm sech}\!\left[\left(\frac{\mathcal R}{\mathcal R_{\rm max}}\right)^2\right].
\label{eq::f_cut}
\end{equation}

To compute the current, we assume that particles stream along a flux tube aligned with the background magnetic field, with longitudinal extent of order the field coherence length \(\lambda_B\). The transverse size of the escaping flux is taken to be the larger of the source size \(R_S\) and the particle Larmor radius in the background field, \(R_L(\mathcal R)\).

In this picture, different nuclei contribute to the instability according to the electric current they carry at fixed rigidity. The current associated with particles with rigidity larger than \(\mathcal R\) is then
\begin{align}
J_{\rm CR}(>\mathcal R)=\sum_{(A,Z)}
\left[
Z e \int_{\mathcal R}^{\mathcal R_{\rm max}} d\mathcal R'\,
\frac{q_{\rm acc}^{(A,Z)}(Z\mathcal R')}
{\pi\left[R_S+R_L(\mathcal R')\right]^2}
\right]
 \nonumber \\
\simeq \frac{e g_Z L}{\Lambda\,\pi}\,
\frac{1}{\mathcal R\left[R_S+R_L(\mathcal R)\right]^2},
\label{eq::current_with_nuclei}
\end{align}
where we introduced the composition-dependent factor
\begin{equation}
g_Z = \sum_{(A,Z)} \frac{\eta_{(A,Z)}}{Z}.
\label{eq::charge_modification}
\end{equation}

In Eq.~\eqref{eq::current_with_nuclei}, we have assumed that each nucleus contributes to the electric current as $Z$ protons with the same rigidity.
It is thus convenient to define an effective luminosity,
\begin{equation}
\mathcal L \equiv g_Z L,
\end{equation}
which is the quantity that actually controls the growth of the instability, and depends on the mass fractions of the accelerated nuclei.

Following~\citet{Cermenati2026aa}, we assume that the magnetic perturbations grow until the amplified field reaches the saturation level
\begin{equation}
\delta B = \sqrt{\frac{4\mathcal L}{\Lambda c R_S^2}}
\approx
25\,{\rm nG}\,
\left(\frac{\mathcal L}{10^{44}\,{\rm erg\,s^{-1}}}\right)^{1/2}
\left(\frac{R_S}{0.3\,{\rm Mpc}}\right)^{-1},
\label{eq::amplified_field}
\end{equation}
where the numerical estimate refers to our fiducial parameters.

At saturation, the fastest-growing mode reaches a scale comparable to the Larmor radius of the particles carrying the current in the amplified field:
\begin{equation}
k_{\rm max}(\mathcal R)=\frac{4\pi}{c}\frac{J_{\rm CR}(>\mathcal R)}{\delta B}
=\frac{1}{R_L(\mathcal R)}
=\frac{e\delta B}{\mathcal R}.
\label{eq::kappa_max}
\end{equation}
This condition identifies the scale at which resonant scattering becomes effective.

The corresponding saturation time is
\begin{equation}
\tau_{\rm sat}(\mathcal R)\approx
1\,{\rm Gyr}\,
\left(\frac{\mathcal R}{0.1\,{\rm EV}}\right)
\left(\frac{\mathcal L}{10^{44}\,{\rm erg\,s^{-1}}}\right)^{-1}
\left(\frac{R_S}{0.3\,{\rm Mpc}}\right)^2
\left(\frac{\rho}{\rho_f}\right)^{1/2},
\label{eq::saturation_time}
\end{equation}
where \(\rho_f \simeq \delta_f \rho_b\), with \(\rho_b \simeq 4.2\times10^{-31}\,{\rm g\,cm^{-3}}\) the mean cosmic baryon density, and we adopt as a reference an overdensity \(\delta_f \simeq 30\), representative of the warm-hot intergalactic medium (WHIM) in cosmic filaments~\citep{Zhang2024mnras}.

To assess whether the instability has enough time to develop, we assume that sources remain active over cosmological timescales. In this simplified picture, the age of a source observed at redshift \(z\) is taken to be
\begin{equation}
T_{\rm age}(z)=\int_{z}^{z_{\rm max}} dz'\,
\left|\frac{dt}{dz'}\right|,
\label{eq::Source_Age}
\end{equation}
where \(|dt/dz| = H(z)^{-1}/(1+z)\), and we take \(z_{\rm max}\sim 10\) to mark the onset of any possible UHECR source.
The adopted cosmology is the Planck $\Lambda$CDM model, with h = 0.67, $\Omega_m$ = 0.32, and $\Omega_\Lambda$ = 0.68~\citep{Planck2016aa}.

This timescale is especially relevant for the confined component, because the density of particles trapped in the self-generated turbulence, and hence the associated production of secondaries, accumulates over the entire source lifetime.

We then define a critical rigidity \(\mathcal R_c=\mathcal R_c(z)\) through the condition
\begin{equation}
\tau_{\rm sat}(\mathcal R_c)=T_{\rm age}(z).
\end{equation}
For \(\mathcal R \lesssim \mathcal R_c\), the instability has enough time to reach resonant scales, and particle transport becomes diffusive in the amplified turbulence. Following~\citet{Cermenati2026aa}, we model the diffusion coefficient as
\begin{equation}
D(\mathcal R)=\frac{c}{3}\frac{\mathcal R_c}{e\delta B}
\left[
\left(\frac{\mathcal R}{\mathcal R_c}\right)
+
\left(\frac{\mathcal R}{\mathcal R_c}\right)^2
\right].
\label{eq::diffusion_coefficient}
\end{equation}
This coefficient acquires a redshift dependence through \( \mathcal R_c(z) \). This expression interpolates between a Bohm-like regime at low rigidity and a faster, quasi-ballistic scaling at high rigidity~\citep{Mollerach2013jcap}.

Once diffusion is established, the cosmic-ray distribution develops a pressure gradient along the magnetic field. The plasma is then expected to respond with a bulk motion of the order of the Alfv\'en speed in the amplified field (see also~\citealt{Blasi2019prl}):
\begin{equation}
V_A = \frac{\delta B}{\sqrt{4 \pi \rho}}
\approx
0.02 \,\frac{{\rm Mpc}}{{\rm Gyr}}\,
\left(\frac{\mathcal L}{10^{44} \, {\rm erg\,s^{-1}}}\right)^{1/2}
\left(\frac{R_S}{0.3 \, {\rm Mpc}}\right)^{-1}
\left(\frac{\rho}{\rho_f}\right)^{-1/2},
\end{equation}

This speed may therefore inherit a redshift dependence through the ambient density. As discussed in~\citet{Cermenati2026aa}, if the luminosity is sufficiently large, plasma advection may remove particles faster than the instability can fully develop, thereby weakening self-confinement. This effect translates into a luminosity-dependent lower bound on \(\lambda_B\) for self-confinement to operate. As we will show below, for our reference parameters this condition is not restrictive, since the required luminosities are well above those expected for typical UHECR sources.

With the diffusion coefficient and advection speed in place, we can estimate the characteristic times required for a particle of rigidity \(\mathcal R\) to cross a distance \(\lambda_B\), either by diffusion in the self-generated turbulence or by advection with the background plasma:
\begin{equation}
\tau_{\rm diff}(\mathcal R)=\frac{\lambda_B^2}{4D(\mathcal R)},
\qquad
\tau_{\rm adv}=\frac{\lambda_B}{V_A}.
\end{equation}

Below a critical rigidity \( \mathcal R_{\rm cut} \), the transport time across one coherence length becomes comparable to the age of the source and the escaping flux will be suppressed.
In the case of diffusion-dominated transport, this critical rigidity corresponds to
\begin{equation}
\mathcal R_{\rm cut}\approx
1\,{\rm EV}\,
\left(\frac{\lambda_B}{30\,{\rm Mpc}}\right)
\left(\frac{\mathcal L}{10^{44}\,{\rm erg\,s^{-1}}}\right)^{3/4}
\left(\frac{R_S}{0.3\,{\rm Mpc}}\right)^{-3/2}
\left(\frac{\rho}{\rho_f}\right)^{-1/4}.
\label{eq::rigidity_cutoff}
\end{equation}

Equation~\eqref{eq::rigidity_cutoff} makes explicit that the characteristic suppression scale is set by the interplay between current-driven magnetic amplification (through $L/R_S^2$) and the density and magnetic structure of the surrounding medium.

\subsection{UHECR confinement and interactions}
\label{sec:confinementandinteractions}

A complete description of particle transport inside the confinement region would require a dedicated numerical treatment, including energy losses, nuclear disintegration, and the coupled time evolution of particles and self-generated waves. In particular, the growth of the magnetic perturbations depends on the cosmic-ray current, while the current itself is controlled by transport. As discussed in~\citet{Cermenati2026aa}, the propagation is expected to evolve from an initial ballistic phase to a diffusive regime on a timescale shorter than the formal saturation time, because lower-rigidity particles first generate turbulence on scales smaller than the gyroradius of the highest-energy particles. At the same time, the diffusion coefficient evolves as the magnetic fluctuations grow. A full treatment of this non-linear problem is numerically demanding and lies beyond the scope of the present work. However, these complications are not expected to qualitatively modify the physical picture relevant for our discussion.

We therefore adopt a phenomenological approach and describe the UHECR population confined within the turbulent region (around the source) using a Leaky Box model. This framework is analogous to that commonly employed for cosmic-ray transport in the Galactic halo and in source-environment confinement scenarios~\citep{Unger2015prd,Muzio2019prd,Muzio2022prd}. Our aim is not to reproduce the detailed time-dependent evolution of the system, but rather to capture, in a transparent way, the competition between escape and interactions in the self-confined region.

In the Leaky Box picture, the equilibrium density of each nuclear species is determined by the balance between injection, escape, and interactions. The escape time includes both diffusive transport through the self-generated turbulence and advection with the background plasma, and can be written as
\begin{equation}
\tau_{\rm esc}^{(A,Z)}(E,z)=
\left[
\frac{1}{\tau_{\rm adv}(z)}+\frac{1}{\tau_{\rm diff}^{(A,Z)}(E,z)}
\right]^{-1}
=
\left[
\frac{V_A(z)}{\lambda_B}+\frac{4D^{(A,Z)}(E,z)}{\lambda_B^2}
\right]^{-1},
\label{eq::escape_time}
\end{equation}
where
\[
D^{(A,Z)}(E,z)\equiv D(E/Z,z),
\qquad
\tau_{\rm diff}^{(A,Z)}(E,z)\equiv \frac{\lambda_B^2}{4D^{(A,Z)}(E,z)}.
\]
Here we have made explicit the dependence on the source redshift.
The relevant interaction channels depend on the nuclear species. For protons, the dominant processes are Bethe--Heitler pair production and photopion production, whereas for nuclei the main channel is photodisintegration~\citep{Boncioli:2023gbl}. In the latter case, the Lorentz factor of the fragments is approximately conserved, so the process can be treated effectively as a decay in mass number~\citet{Aloisio2013aph}:
\begin{equation}
\tau_{\rm int}^{(A,Z)}(E,z)\equiv \tau_{\gamma N}^{(A,Z)}(E,z)=
\left|\frac{1}{A}\frac{dA}{dt}\right|^{-1}.
\label{eq::tau_int_nuclei}
\end{equation}
The dominant contribution typically comes from the Giant Dipole Resonance, which mostly leads to the emission of a single nucleon. Since emitted neutrons decay rapidly on the scales of interest, we do not distinguish between proton and neutron emission in the subsequent treatment. We compute the corresponding interaction rate as in Eq.~2 of~\citet{Aloisio2013aph}, adopting the cross-section parameterization of~\citet{Puget1976apj}. As target photon fields, we include both the CMB and the adopted EBL model~\cite{Gilmore2012mnras}.

\begin{figure*}[t]
\centering
\includegraphics[width=0.4\linewidth]{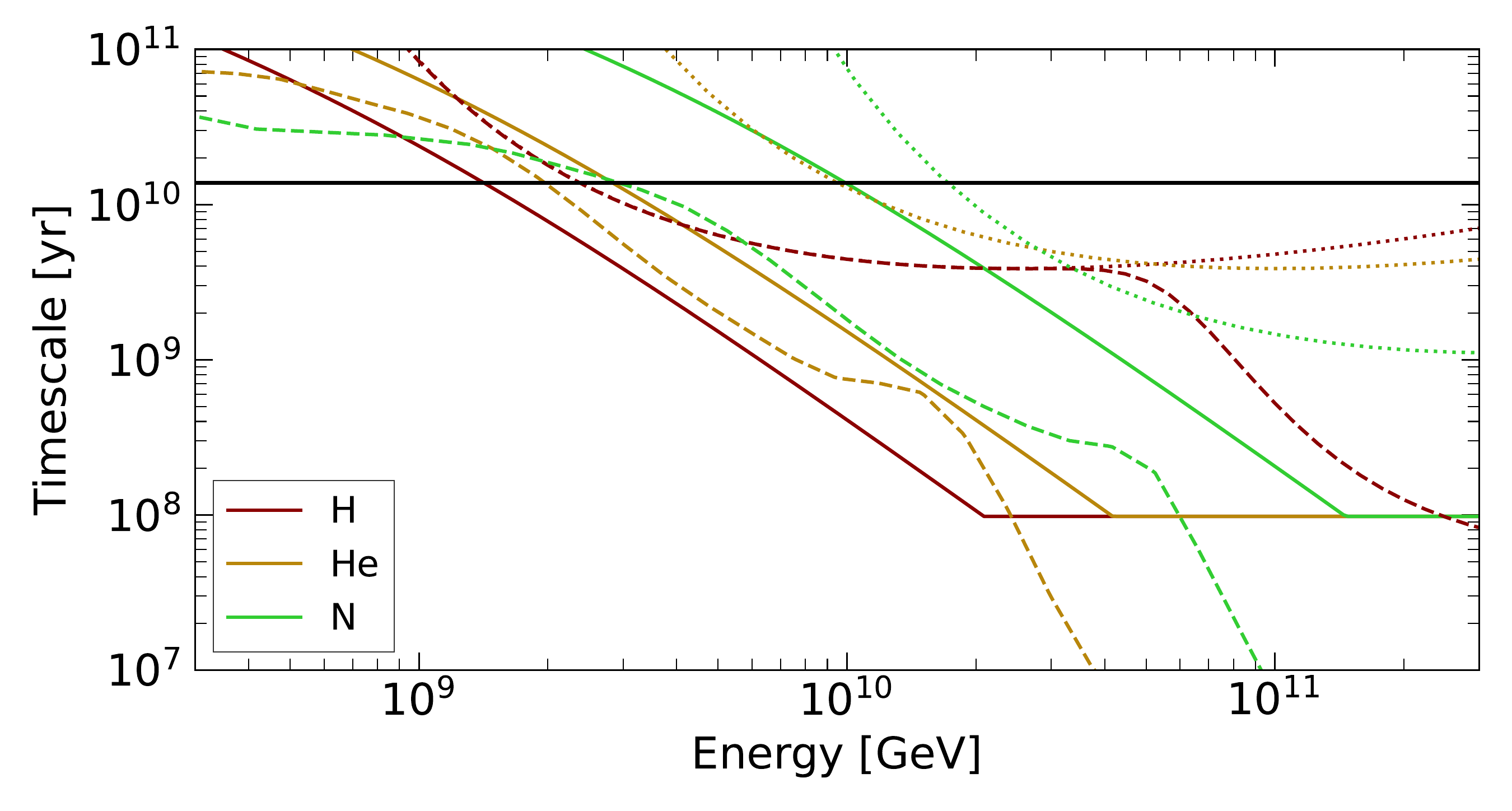}
\includegraphics[width=0.4\linewidth]{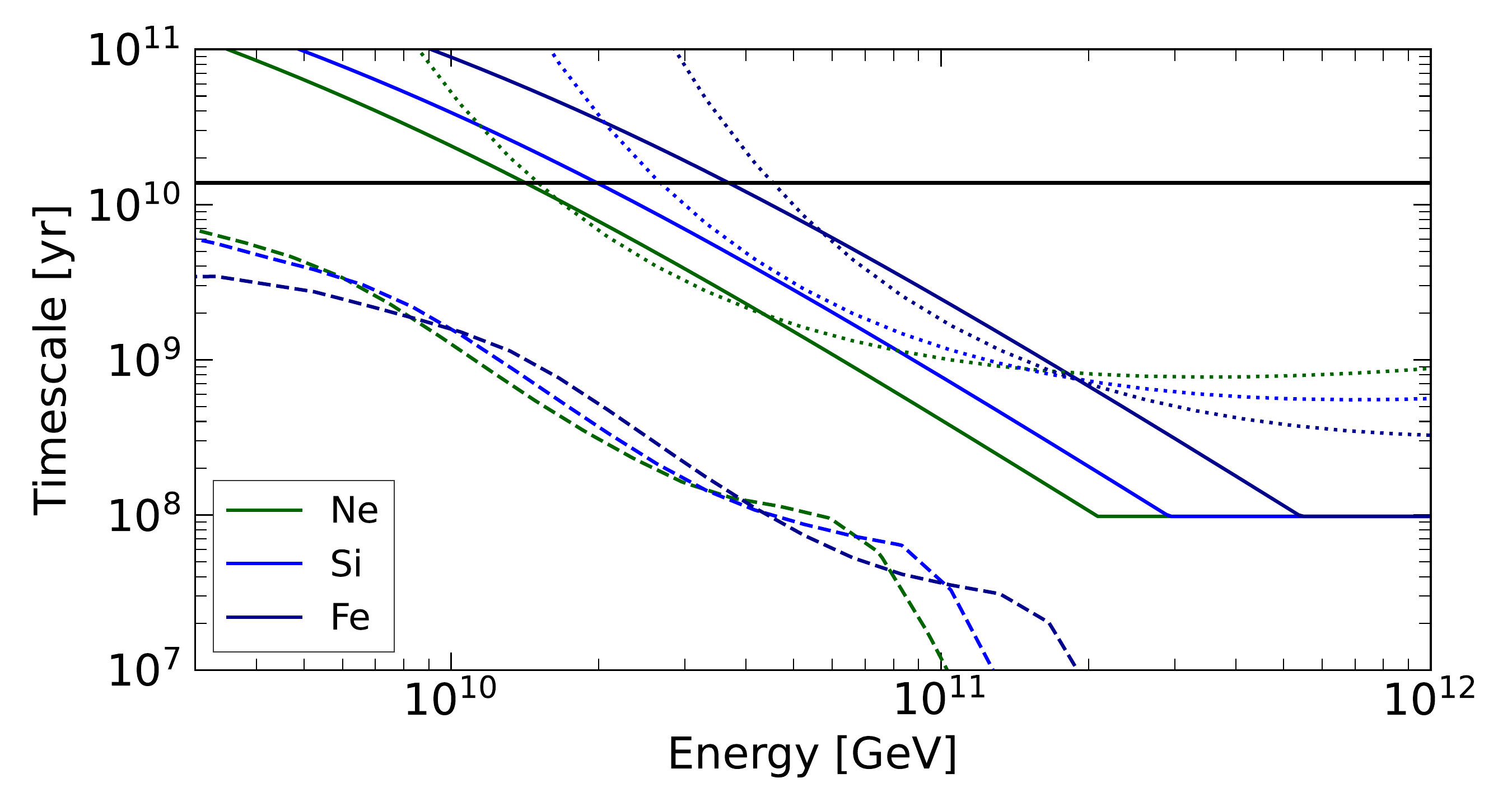}
\caption{Relevant timescales (computed at \(z=0\)) for representative light (left panel) and heavy (right panel) nuclei in the reference scenario. Solid lines show the escape time, including the lower bound set by the flux-tube crossing time, \(\lambda_B/c \sim 0.1\)~Gyr. Dashed lines indicate the relevant interaction timescales (\(\gamma N\) for nuclei and \(\gamma p\) for protons), while dotted lines show the timescale for Bethe--Heitler pair production. The black horizontal line marks the Hubble time, \(H_0^{-1}\).}
\label{fig::timescales}
\end{figure*}

In addition to photodisintegration, nuclei also undergo Bethe--Heitler pair production. Since this process changes the particle energy but not the nuclear species, we treat it separately from the disintegration chain and show its corresponding timescale explicitly in Fig.~\ref{fig::timescales}.

For protons we can define an effective loss timescale
\begin{equation}
\tau_{\rm int}^{(1,1)}(E,z)\equiv \tau_{\gamma p}(E,z)=
\left|\frac{db(E,z)}{dE}\right|^{-1},
\label{eq::tau_int_protons}
\end{equation}
where \(b(E,z)\) is the proton energy-loss function, computed as described in~\citet{Berezinsky2006prd}.

Figure~\ref{fig::timescales} compares the escape time in our reference scenario with the relevant interaction timescales for representative light and heavy nuclei. In the fiducial case, photodisintegration is highly efficient for heavy nuclei, with \(\tau_{\rm esc}^{(A,Z)} \gtrsim 10\,\tau_{\gamma N}^{(A,Z)}\). 
This comparison already illustrates the key point of our model: once self-confinement becomes effective, nuclei can remain trapped long enough for interactions with the ambient radiation backgrounds to substantially reshape the escaping composition and enhance the associated production of secondary particles.

Within this framework, the equilibrium density of primary UHECRs in the confinement region satisfies
\begin{equation}
\frac{n_{\rm I}^{(A,Z)}(E,z)}{\tau_{\rm esc}^{(A,Z)}(E,z)}
+
\frac{n_{\rm I}^{(A,Z)}(E,z)}{\tau_{\rm int}^{(A,Z)}(E,z)}
=
q_{\rm acc}^{(A,Z)}(E),
\label{eq::LB_primary}
\end{equation}
whose solution is
\begin{equation}
n_{\rm I}^{(A,Z)}(E,z)=
q_{\rm acc}^{(A,Z)}(E)\,
\tau_{\rm eff}^{(A,Z)}(E,z),
\label{eq::primary_nuclei_density}
\end{equation}
where the effective confinement time is
\begin{equation}
\tau_{\rm eff}^{(A,Z)}(E,z)=
\frac{\tau_{\rm esc}^{(A,Z)}(E,z)\,\tau_{\rm int}^{(A,Z)}(E,z)}
{\tau_{\rm esc}^{(A,Z)}(E,z)+\tau_{\rm int}^{(A,Z)}(E,z)}.
\label{eq::EffectiveTimescale}
\end{equation}

The same formalism can be extended to secondary nuclei of mass \(A\), produced by a primary nucleus of mass \(A^{\rm I}\) through a chain of photodisintegration processes (see also~\citealt{Unger2015prd}). Their steady-state density can be written as
\begin{equation}
n_{\rm II}^{(A,Z)}(E,z)=
q_{\rm acc}^{(A^{\rm I},Z^{\rm I})}\!\left(\frac{A^{\rm I}}{A}E\right)
\frac{A^{\rm I}}{A}
\prod_{A'=A+1}^{A^{\rm I}}
\left[
\frac{\tau_{\rm eff}^{(A',Z')}\!\left(\frac{A'}{A}E,z\right)}
{\tau_{\rm int}^{(A',Z')}\!\left(\frac{A'}{A}E,z\right)}
\right]
\tau_{\rm eff}^{(A,Z)}(E,z).
\label{eq::Secondary_Nuclei_density}
\end{equation}
This expression makes explicit that the abundance of a secondary fragment is determined by the cumulative probability of surviving each step of the disintegration chain while remaining confined in the source environment.

In addition, each photodisintegration step releases a nucleon. Under our approximation, neutrons decay rapidly and are therefore counted as protons. The total equilibrium density of confined protons is thus given by the sum of a primary and a secondary contribution:
\begin{equation}
\begin{split}
n_{\rm I}^{\rm (1,1)}(E,z) &= q_{\rm acc}^{\rm H}(E)\,\tau_{\rm eff}^{\rm H}(E,z), \\
n_{\rm II}^{\rm (1,1)}(E,z) &= q_{\rm acc}^{(A^{\rm I},Z^{\rm I})}(A^{\rm I}E)\,A^{\rm I}
\sum_{A'=2}^{A^{\rm I}}
\left[
\prod_{A''=A'}^{A^{\rm I}}
\frac{\tau_{\rm eff}^{(A'',Z'')}(A''E,z)}
{\tau_{\rm int}^{(A'',Z'')}(A''E,z)}
\right]
\tau_{\rm eff}^{\rm (1,1)}(E,z).
\end{split}
\label{eq::Protons_density}
\end{equation}

The solutions above describe the equilibrium densities of particles confined within the self-generated turbulent region. However, the escaping flux is not determined solely by these densities. 
In particular, at low rigidity, where the escape time becomes larger than the age of the source, particles do not reach the boundary of the coherence domain, preventing the system from reaching its equilibrium configuration within the source lifetime.
This effect is not captured by the Leaky Box treatment alone and must therefore be included separately when computing the escaping spectrum.

The Leaky Box model provides only an effective description of transport on scales comparable to, or smaller than, the diffusive length,
\begin{equation}
l_D(\mathcal R,z)=\sqrt{4D(\mathcal R,z)\,T_{\rm age}(z)},
\label{eq::diffusive_length}
\end{equation}
which represents the typical distance explored by particles over the source lifetime. This quantity therefore sets the characteristic size of the region over which the Leaky Box approximation can be regarded as representative of the average confinement volume.

In our framework, the low-rigidity suppression, needed to reproduce UHECR observations, arises because particles cannot travel across one magnetic-field coherence length within the available source lifetime. In other words, for rigidities below \(\mathcal R_{\rm cut}\), the diffusive length becomes smaller than the coherence length, \(l_D<\lambda_B\). As a consequence, particles cannot reach the boundary of the confinement region, and the escaping flux is suppressed.

To account for the fact that particles with \(\mathcal R<\mathcal R_{\rm cut}\) can propagate only over distances smaller than \(\lambda_B\), we introduce an exponential attenuation factor inspired by the Green function of the one-dimensional advection--diffusion equation (see Appendix~\ref{apx::AdvectionDiffusion}):
\begin{equation}
\mathcal G(\mathcal R,z)=
\exp\left[
-\frac{\bigl(\lambda_B-V_A(z)\,T_{\rm age}(z)\bigr)^2}{4D(\mathcal R,z)\,T_{\rm age}(z)}
\right].
\label{eq::cutoff_GreenFunction}
\end{equation}
In this way, the position of the low-rigidity suppression is directly linked to a physical property of the source environment, namely the magnetic-field coherence length \(\lambda_B\).

We then model the escaping flux as 

\begin{equation}
\phi_{\rm I/II}^{(A,Z)}(E,z)=
\frac{n_{\rm I/II}^{(A,Z)}(E,z)}{\tau_{\rm esc}^{(A,Z)}(E,z)}\,
\mathcal G(E/Z,z).
\label{eq::escaping_flux}
\end{equation}

\begin{figure*}[t]
\centering
\includegraphics[width=0.4\linewidth]{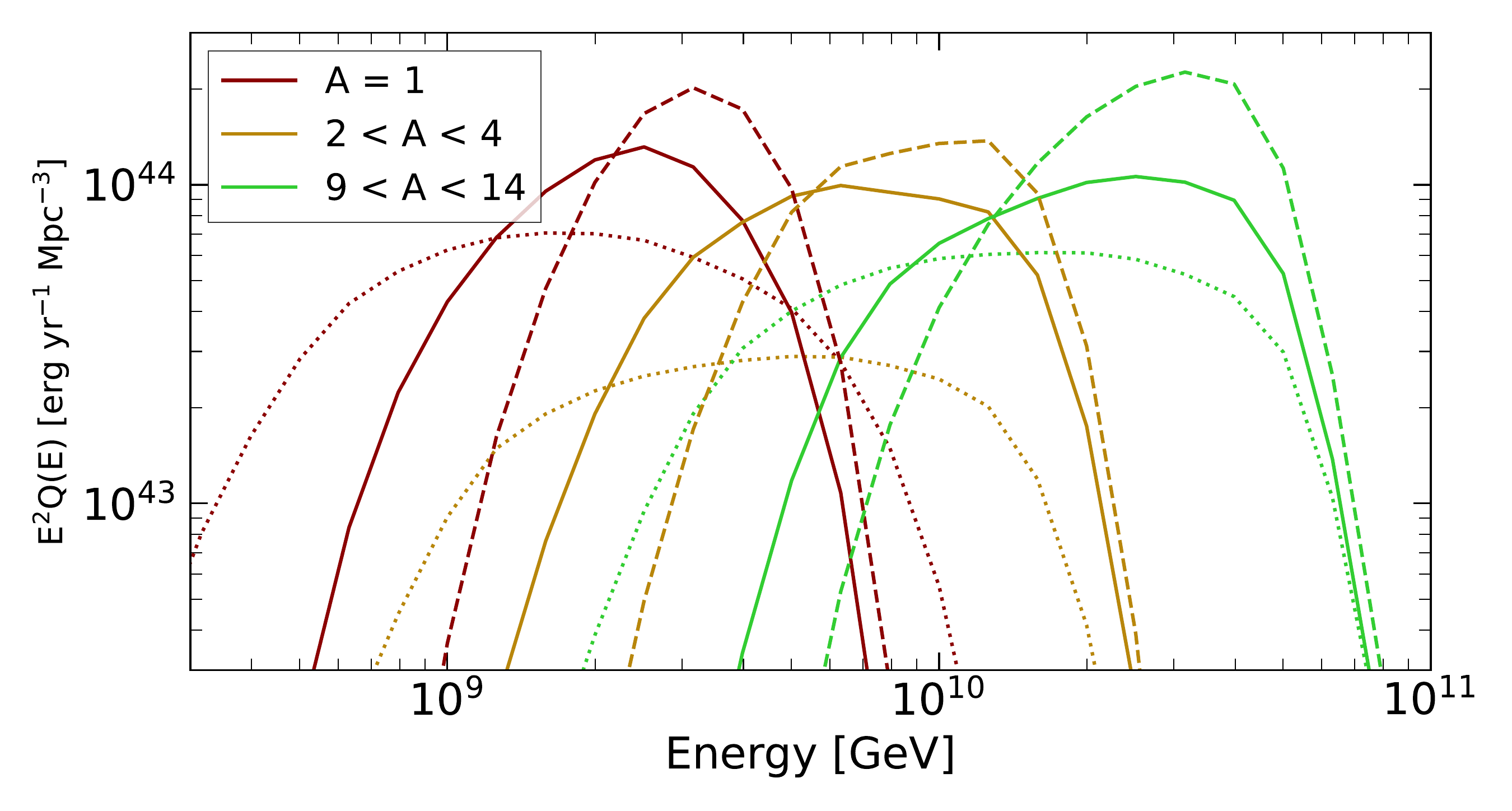}
\includegraphics[width=0.4\linewidth]{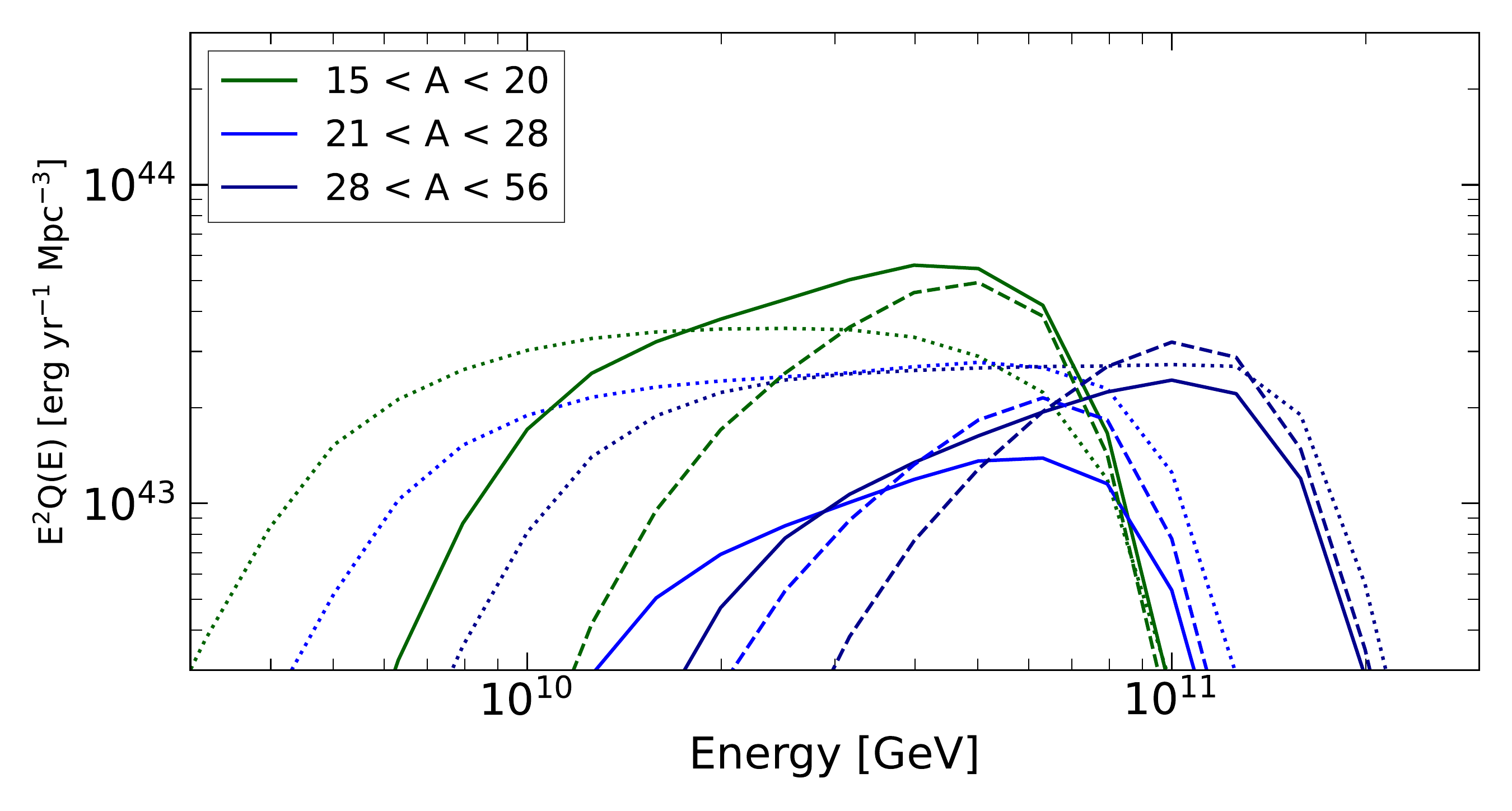}
\caption{Emissivity \(E^2Q(E)\), computed at $z=0$, of different mass groups in the presence of self-confinement, shown for three choices of the magnetic-field coherence length: \(\lambda_B = 15\) Mpc (dotted), \(\lambda_B = 30\) Mpc (solid), and \(\lambda_B = 50\) Mpc (dashed). For each mass interval, the plotted emissivity is the sum of the primary contribution and all secondary nuclei generated by the disintegration of heavier primaries and ending in the same mass group. Each contribution is weighted by the source number density and the luminosity fractions needed to reproduce the observed UHECR data, as reported in Table~\ref{tab::Parameters}.}
\label{fig::emissivity}
\end{figure*}

Finally, we compute the comoving UHECR emissivity from a population of identical sources as
\begin{equation}
\widetilde{Q}_{\rm I/II}^{(A,Z)}(E,z)=
\widetilde{n}_s\,\phi_{\rm I/II}^{(A,Z)}(E,z),
\label{eq::Nuclei_Emissivity}
\end{equation}
where \(\widetilde{n}_s\) is the comoving source number density.

To illustrate the shape of the escaping emissivity, Fig.~\ref{fig::emissivity} shows the resulting emissivities, at $z=0$, of primary and secondary nuclei, grouped into mass intervals, for \(\lambda_B = 15,\ 30,\ 50\) Mpc. The calculations are performed using the parameters listed in Table~\ref{tab::Parameters}, which are chosen to reproduce the main features of the observed UHECR spectrum and composition, as discussed below.

Fig.~\ref{fig::emissivity} already highlights the main effect of the model. As the magnetic-field coherence length increases, the low-rigidity suppression becomes more pronounced, progressively reducing the escape of particles at low energy. As a result, the effective emissivity is shifted toward the highest energies accessible at the source and can exhibit an apparent hardening just below the cutoff energy. In this sense, the confinement effect acts as a spectral filter, reshaping the spectrum emerging from the source environment even when the injected spectrum itself remains unchanged.

Once the emissivities of the primary and secondary nuclei emerging from the confinement region have been determined (see Eq.~\eqref{eq::Nuclei_Emissivity}), we compute the propagated spectrum and composition at Earth using the \texttt{SimProp} code~\citep{Aloisio2017jcap}. Since this is a standard procedure, we defer a detailed description of the calculation of the spectrum and composition at Earth to Appendix~\ref{apx::spectrum_and_composition}.

As shown by \citet{Cermenati2026aa}, the spectral shape below the suppression scale depends not only on confinement, but also on the luminosity distribution of the sources, whereas the position of the suppression is determined mainly by the peak luminosity of that distribution. To remove this degeneracy, here we neglect any luminosity distribution and approximate the source population with identical sources whose luminosity is set equal to the peak of the distribution.

While this simplification affects the detailed spectral shape below the suppression scale and produces sharper features in both the energy spectrum and the evolution of \(X_{\rm max}\), it provides a cleaner framework to identify the region of astrophysical parameter space in which confinement is compatible with current UHECR data.

\subsection{Cosmogenic neutrinos and gamma-rays}

Particles that remain confined within magnetized regions for timescales longer than the source lifetime can continue to interact with the ambient background radiation fields, thereby producing cosmogenic neutrinos and gamma-rays.

In our setup, the relatively low maximum rigidity strongly suppresses photopion production on the CMB because of its high energy threshold (see Fig.~\ref{fig::timescales}). For protons, photopion production on the EBL is still active and therefore contributes to the cosmogenic neutrino flux, although its impact on the parent proton spectrum remains modest because the corresponding interaction timescale is typically longer than the age of the Universe. For nuclei, instead, photodisintegration remains the dominant process in the energy range where photopion production on the EBL could become relevant. As a result, the main contribution of nuclei to the cosmogenic secondary flux is expected to arise indirectly, through the secondary protons produced by photodisintegration during confinement.

By definition, confined particles have an escape time longer than the source age. Therefore, in analogy with Section~2.2, we describe their equilibrium density as
\begin{equation}
    n_{\rm I/II}^{\rm (1,1)\star}(E,z) = n^{\rm (1,1)}_{\rm I/II}(E,z)
    \left[1 - \exp\left(- \frac{T_{\rm age}(z)}{\tau_{\rm eff}(E,z)} \right)\right].
\label{eq::Protons_density_confined}
\end{equation}
This expression smoothly interpolates between the limit in which both escape and interaction timescales exceed the source age, so that particles accumulate over the full source lifetime, and the opposite regime in which interactions deplete the confined population. 

The comoving emissivity of cosmogenic neutrinos produced by confined particles can then be written as the source number density times the neutrino emissivity of a single source,
\begin{equation}
    \widetilde{Q}_\nu (E_\nu,z) =
    \widetilde{n}_s
    \int_{E_\nu}^{+\infty} \frac{dE_p}{E_p}
    \, n^{\rm H\star}_{\rm I+II}(E_p,z)
    \, R_\pi^\nu(E_p,E_\nu,z).
\label{eq::Neutrino_emissivity}
\end{equation}
Here, \(R_\pi^\nu\) [s\(^{-1}\)] is the differential neutrino production rate through photopion interactions, summed over all neutrino flavours~\citep{Kelner2008prd} (see Appendix~\ref{apx::secondaries_released}).

The treatment of gamma-rays is slightly different. Once high-energy photons, electrons, or positrons are produced, they rapidly initiate an electromagnetic cascade~\citep{Berezinsky2016prd}, which does not necessarily remain confined within the magnetized region. In this case, the relevant quantity is the total power injected into the cascade in the form of photons and pairs, whereas the emergent spectrum is largely determined by the kinematics of inverse Compton scattering and photon--photon pair production.

The power injected into the electromagnetic cascade through photopion production can be written as
\begin{equation}
    \Omega_{\pi}(z) = \sum_{i =e^+,e^-, \gamma}
    \int_0^{+\infty} dE_i \, E_i
    \int_{E_i}^{+\infty} \frac{dE_p}{E_p}
    \, n^{\rm H\star}_{\rm I+II}(E_p,z)
    \, R_\pi^{i}(E_p,E_i,z),
\label{eq::Gamma_Pion_EnergyDensity}
\end{equation}
where the index $i$ runs over electrons, positrons, and gamma-rays generated in the photopion production process, and \(R_\pi^{i}\) is the associated differential production rate~\citep{Kelner2008prd}.

An even more important channel for electromagnetic cascades is Bethe--Heitler pair production on background photons, in which energy is transferred from UHE protons to electron--positron pairs.
The corresponding injected power is
\begin{equation}
    \Omega_{ee}(z) =
    \int_0^{+\infty} dE_p \,
    n^{\rm H\star}_{\rm I+II}(E_p,z)\,
    b(E_p,z),
\label{eq::Gamma_Pair_EnergyDensity}
\end{equation}
where \(b(E_p,z)\) is the proton energy-loss rate due to pair production, computed following~\citep{Berezinsky2006prd, Chodorowski1992apj}.

Under the universality approximation for electromagnetic cascades~\citep{Berezinsky2016prd}, the comoving gamma-ray emissivity can then be expressed as
\begin{equation}
\begin{split}
    \widetilde{Q}_\gamma (E_\gamma, z) =
    \, A
    & \times
    \begin{cases}
        \left( \dfrac{E_\gamma}{\epsilon_X(z)} \right)^{-3/2},
        & E_\gamma \leq \epsilon_X(z), \\[0.3cm]
        \left( \dfrac{E_\gamma}{\epsilon_X(z)} \right)^{-2},
        & \epsilon_X(z) < E_\gamma \leq \epsilon_{\rm EBL}^{\rm PP}(z), \\[0.3cm]
        0,
        & E_\gamma > \epsilon_{\rm EBL}^{\rm PP}(z).
    \end{cases}
\end{split}
\label{eq::GammaRay_Emissivity}
\end{equation}
where
\[
A = \widetilde{n}_s \,
    \frac{\Omega_\pi (z) + \Omega_{ee}(z)}
    {\epsilon_X^2(z)\left[2+\ln\!\left(\frac{\epsilon_{\rm EBL}^{\rm PP}(z)}{\epsilon_X(z)}\right)\right]}.
\]
Here, \(\epsilon_X\) and \(\epsilon_{\rm EBL}^{\rm PP}\) are the characteristic cascade energies defined in~\citet{Berezinsky2016prd}. As target photon fields, we include both the CMB and the EBL~\citep{Gilmore2012mnras}.

The calculation of the cosmogenic neutrino and gamma-ray spectra at Earth generated by confined protons is presented in Appendix~\ref{apx::secondaries_confined}. In addition, cosmogenic neutrinos and gamma-rays are also produced by particles after they escape and propagate through the intergalactic medium. Their computation follows the formalism of~\citet{Berezinsky2006prd,Berezinsky2016prd} and is described in Appendix~\ref{apx::secondaries_released}.

\section{Results}
\label{sec:results}

\subsection{Spectrum and composition}
In this section, we compare the predictions of our self-confinement scenario with the observed UHECR spectrum and composition. Since the main effect of self-confinement is the suppression of the escaping flux at low rigidity, we explore how the model changes over a representative range of values of the magnetic-field coherence length, $\lambda_B$.
The dependence of the cutoff rigidity on the model parameters is made explicit in Equation~\eqref{eq::rigidity_cutoff}, which also highlights the main degeneracies of this approach.

For concreteness, we focus on three benchmark cases, $\lambda_B = 15$, 30, and 50 Mpc, which correspond to a suppression of the escaping flux below a characteristic rigidity of approximately $R_{\rm c} \simeq 0.5$, 1, and 2 EV, respectively. The adopted values are consistent with the expected inter-source distance~\citep{Bister2024apj} and the size of filaments in the cosmic web~\citep{universe10070287,Zhang2024mnras}.

The corresponding UHECR emissivities are shown in Fig.~\ref{fig::emissivity}. 
Using the procedure described in Section~\ref{sec:confinementandinteractions}, we then compute the propagated flux at Earth for each nuclear species and compare the total flux with the measured all-particle spectrum. From the same model, we also derive the mean and variance of $\ln A$.

To compare these predictions with composition-sensitive observables, we convert the mean and variance of $\ln A$ into the mean and variance of the shower-maximum distribution, $X_{\rm max}$, using the parameterization of~\citet{Evoli2026aph} for two hadronic interaction models, EPOS-LHC-R~\citep{PhysRevC.92.034906} and Sibyll 2.3e~\citep{PhysRevD.102.063002}.

To account for possible systematic uncertainties in the absolute $X_{\rm max}$ scale, we allow for an overall shift in the hadronic-model response, calibrated on the composition at the highest energies ($\gtrsim 10^{19}$ eV), following the approach adopted in previous analyses~\citep{Unger2015prd}.

Finally, in order to reproduce the average $X_{\rm max}$ evolution at lower energies ($\lesssim 5 \times 10^{18}$ eV), we include an additional extragalactic low-energy (LE) component with properties similar to those inferred in the combined fit of~\citet{PierreAuger2023jcapcombinedfitlowenergy}. We model this component as a standard power-law injection with a soft spectral index, $\gamma = 3$, and fit both its maximum rigidity and nuclear composition.

A reasonable agreement with the observed spectrum and composition can be achieved for $\lambda_B = 30$--$50$ Mpc, with a negligible contribution from primary protons. By contrast, such an agreement is not possible when the low-rigidity suppression lies below the EV scale, as in the case $\lambda_B = 15$ Mpc. For such a weak suppression, the heavy nuclei (Ne, Si, and Fe groups) required to reproduce the data above $10^{19}$ eV extend down into the energy range between the ankle and the instep (Fig.~\ref{fig::spectrum15}). As a consequence, the predicted $X_{\rm max}$ is too low, and a positive shift in its absolute scale is required to match the highest-energy data. Moreover, reproducing the $X_{\rm max}$ evolution below the ankle requires a substantial heavy contribution from the LE population, together with a relatively small maximum rigidity, $R_{\rm max}^{\rm LE} = 0.3$ EV.

At the same time, the shorter escape time implies less photodisintegration during confinement and therefore a smaller flux of secondary protons. Their abundance is too low to account for the observed $\sigma(X_{\rm max})$ around $10^{18}$ eV, where the proton fraction is expected to be largest. As a result, a non-negligible fraction of primary protons from the high-energy (HE) population is needed to increase the $X_{\rm max}$ variance above 2 EeV, while below 1 EeV the variance cannot be reproduced if Peters' cycle scaling is assumed for the secondary population.

\begin{figure*}
\centering
\includegraphics[width=0.40\linewidth]{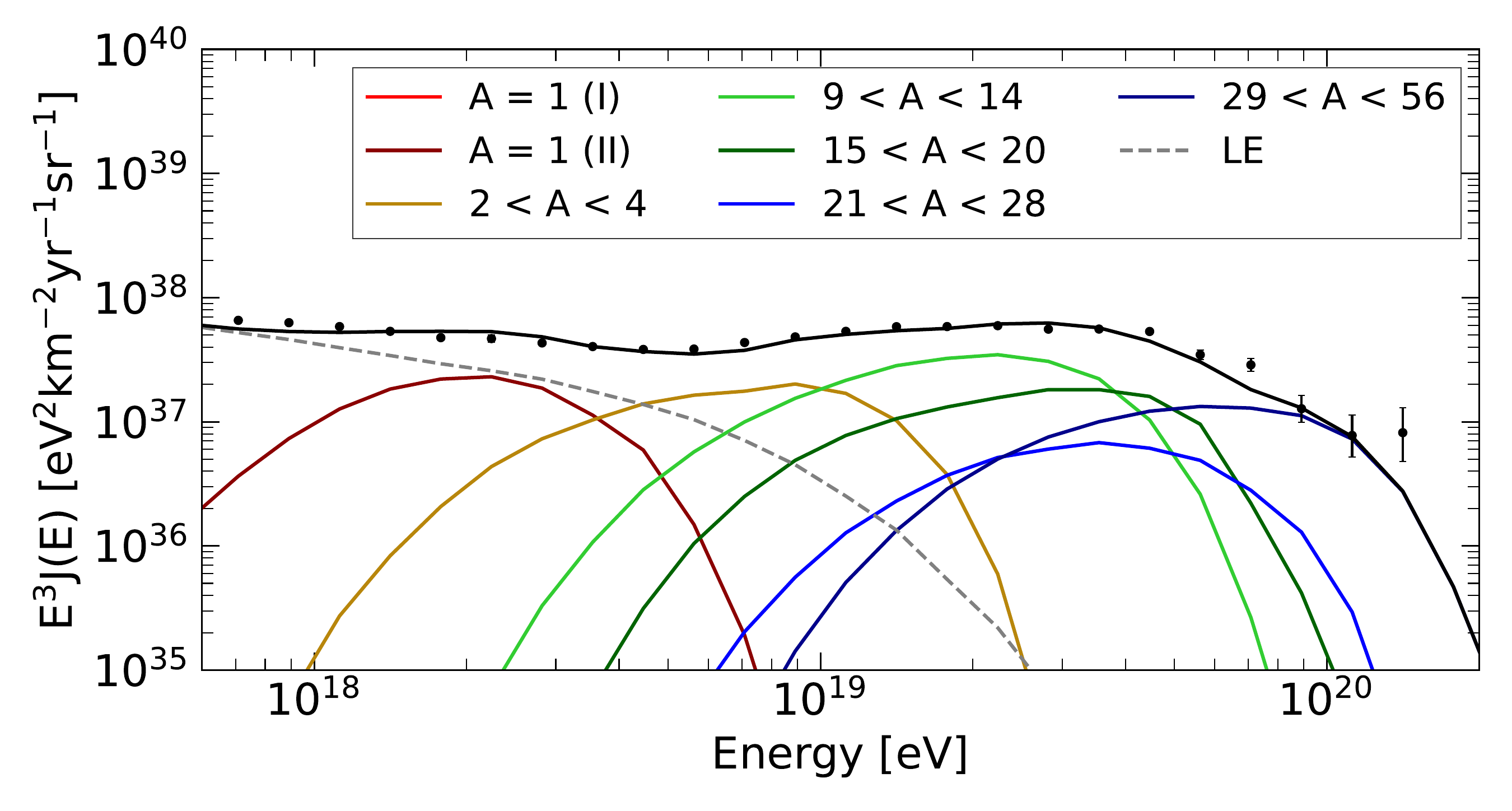}
\includegraphics[width=0.40\linewidth]{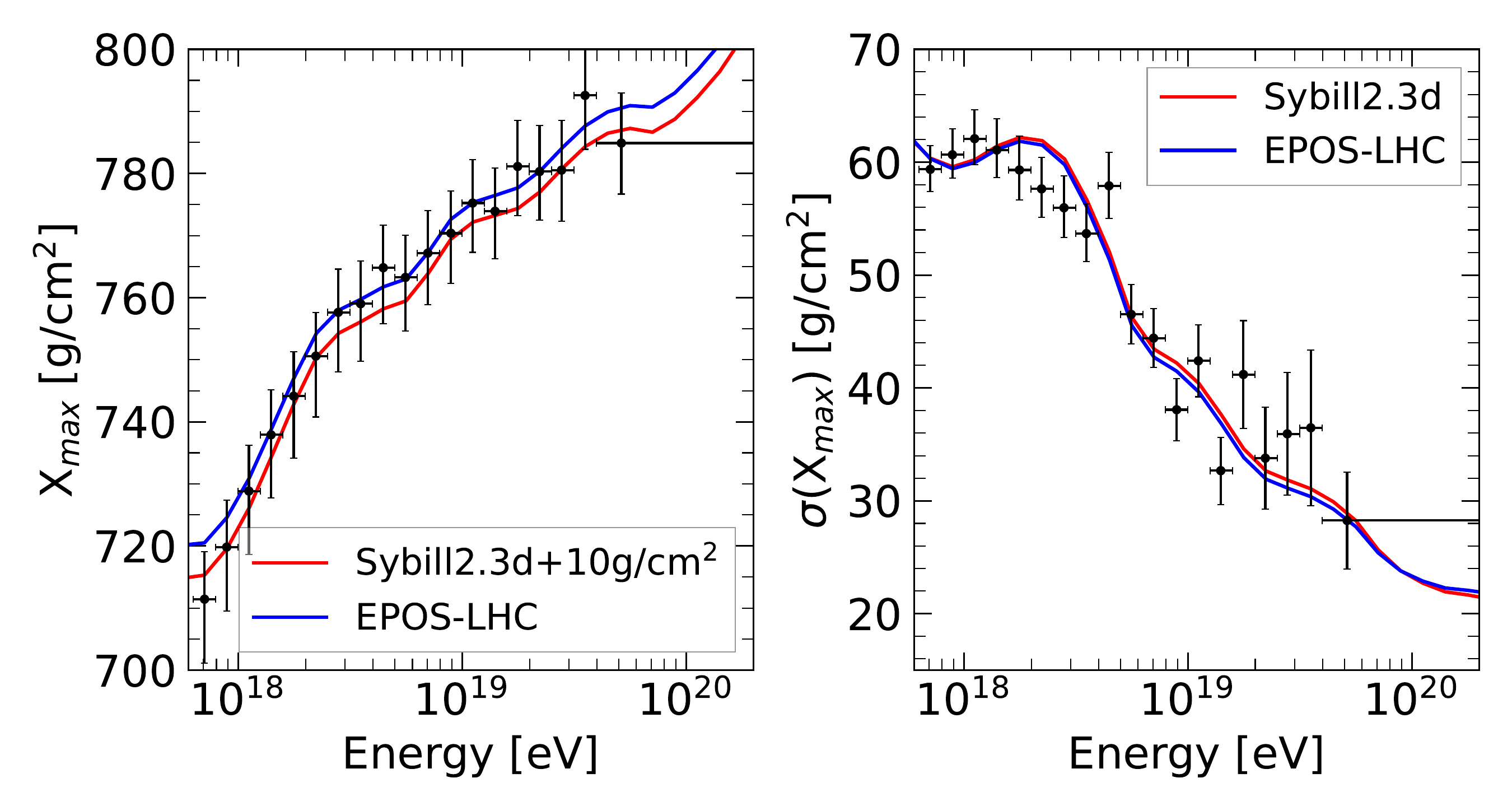}
\caption{\textit{Left}: UHECR spectrum at Earth for $\lambda_B = 30$ Mpc. The spectra of the different mass groups are shown according to the assumed primary composition. The total contribution from the secondary low-energy (LE) population is indicated by the grey dashed line, while the total predicted spectrum is shown by the black solid line. Data points correspond to the all-particle UHECR spectrum measured by Auger~\citep{PierreAuger2020prdspectrum,PierreAuger2021epjcspectrum}. \textit{Right}: Mean and variance of the shower-maximum distribution, derived using the EPOS-LHC and Sibyll parameterizations of~\citet{Evoli2026aph}. The predictions are compared with the Auger $X_{\rm max}$ measurements~\citep{PierreAuger2014prdmassa,PierreAuger2014prdmassb}.}    \label{fig::spectrum30}
\end{figure*}

\begin{figure*}
\centering
\includegraphics[width=0.40\linewidth]{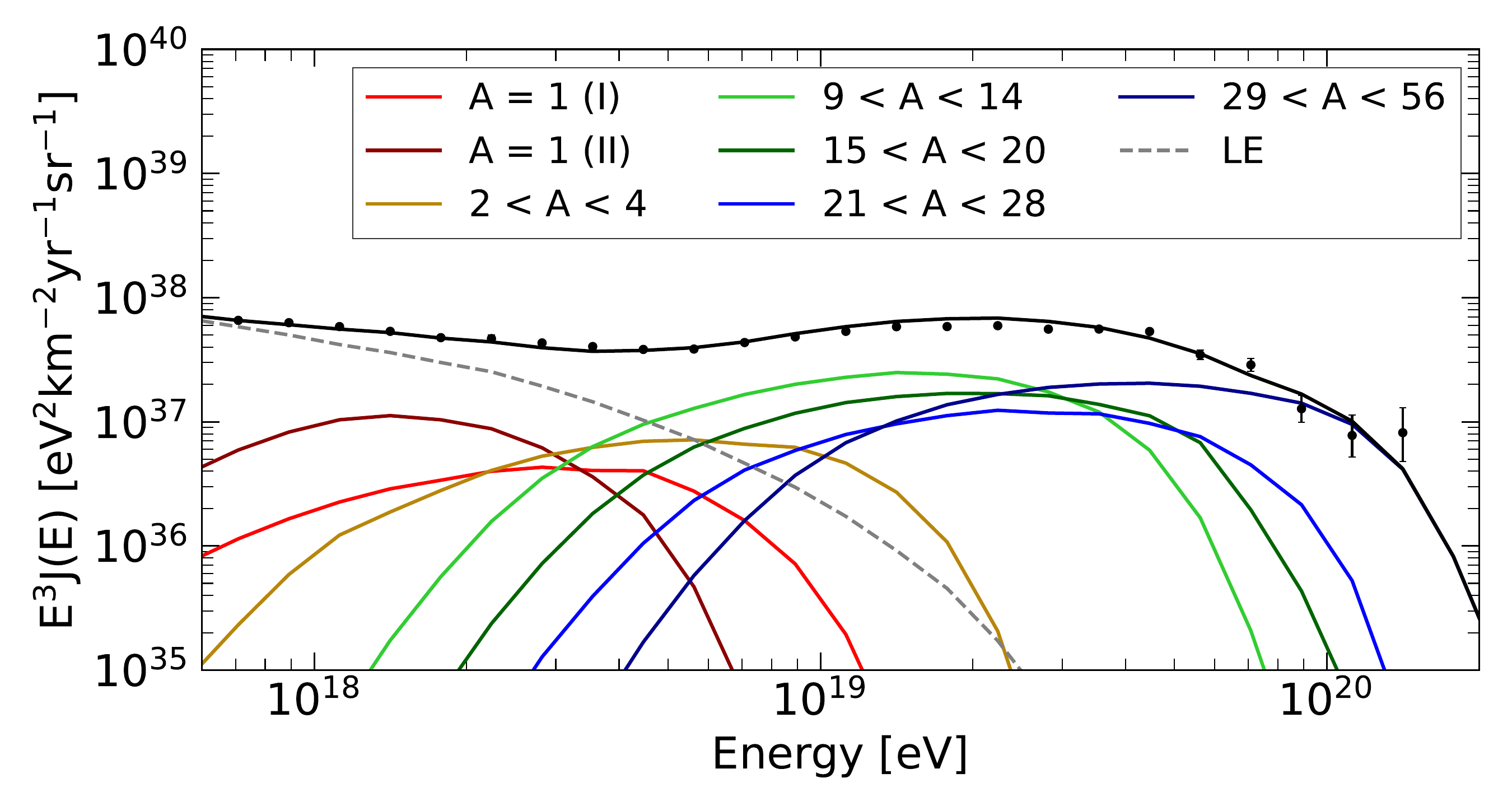}
\includegraphics[width=0.40\linewidth]{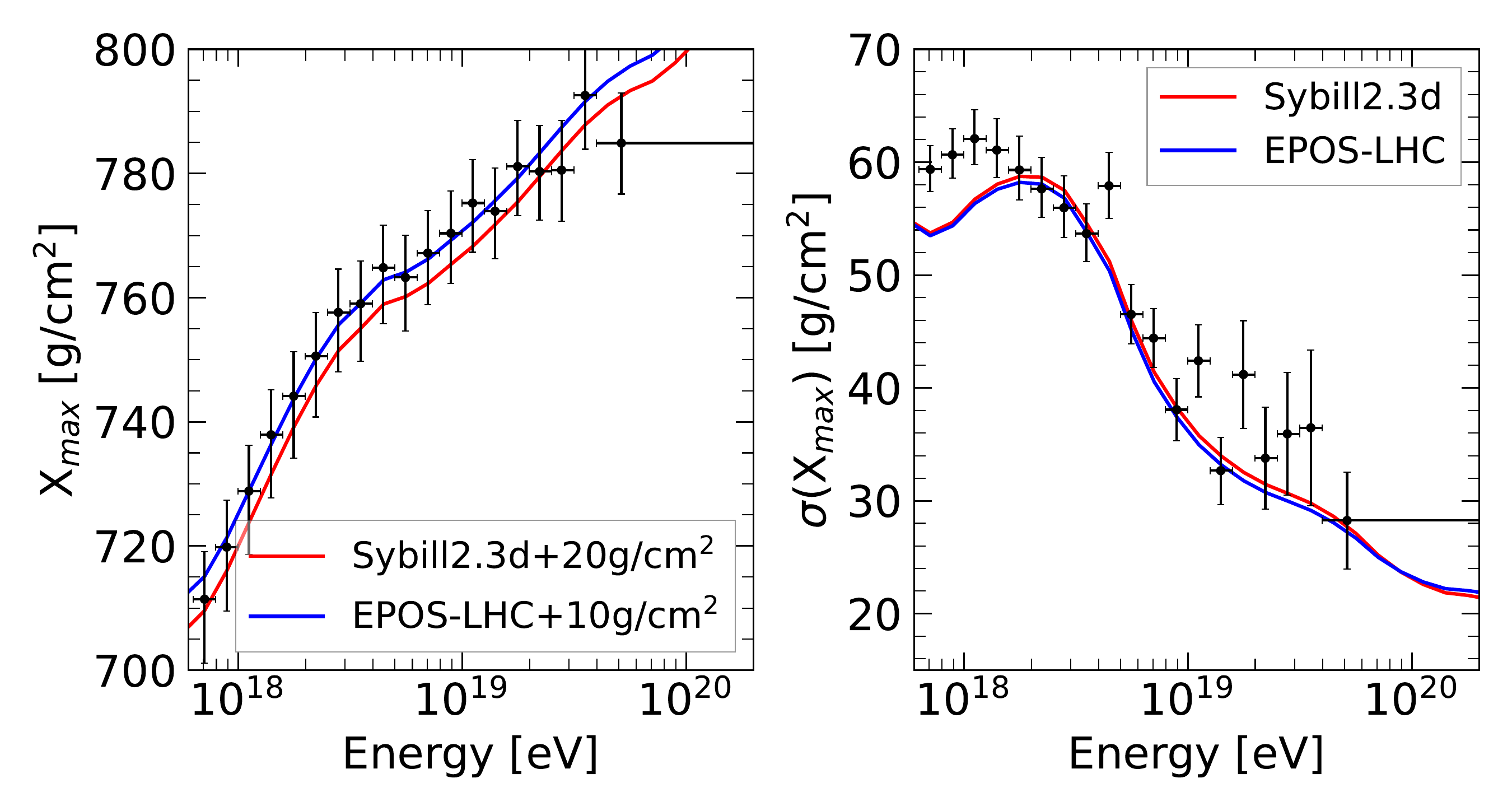}
\caption{UHECR spectrum and corresponding $X_{\rm max}$ moments at Earth, as in Fig.~\ref{fig::spectrum30}, but for $\lambda_B = 15$ Mpc.}    \label{fig::spectrum15}
\end{figure*}

Increasing the low-rigidity suppression to the EV scale, corresponding to $\lambda_B = 30$ Mpc, substantially enhances the production of secondary protons. At the same time, the contribution of heavy nuclei no longer contaminates the ankle--instep region (Fig.~\ref{fig::spectrum30}), leading to larger values of $X_{\rm max}$. EPOS-LHC predictions are fully consistent with observations at energies $\gtrsim 10^{19}$ eV without any shift in the absolute $X_{\rm max}$ scale, while Sibyll still requires a positive shift. In this case, the LE population required below the ankle is also lighter and characterized by a higher maximum rigidity, $R_{\rm max}^{\rm LE} = 0.6$ EV. The predicted variance of $X_{\rm max}$ is consistent with the observations. Around $10^{18}$ eV, the interplay between the maximum rigidity of the secondary LE population and the low-rigidity suppression induced by self-confinement can produce an almost continuous proton spectrum up to the ankle, even if the contribution of primary protons from the self-confined population is negligible.

\begin{figure*}
\centering
\includegraphics[width=0.40\linewidth]{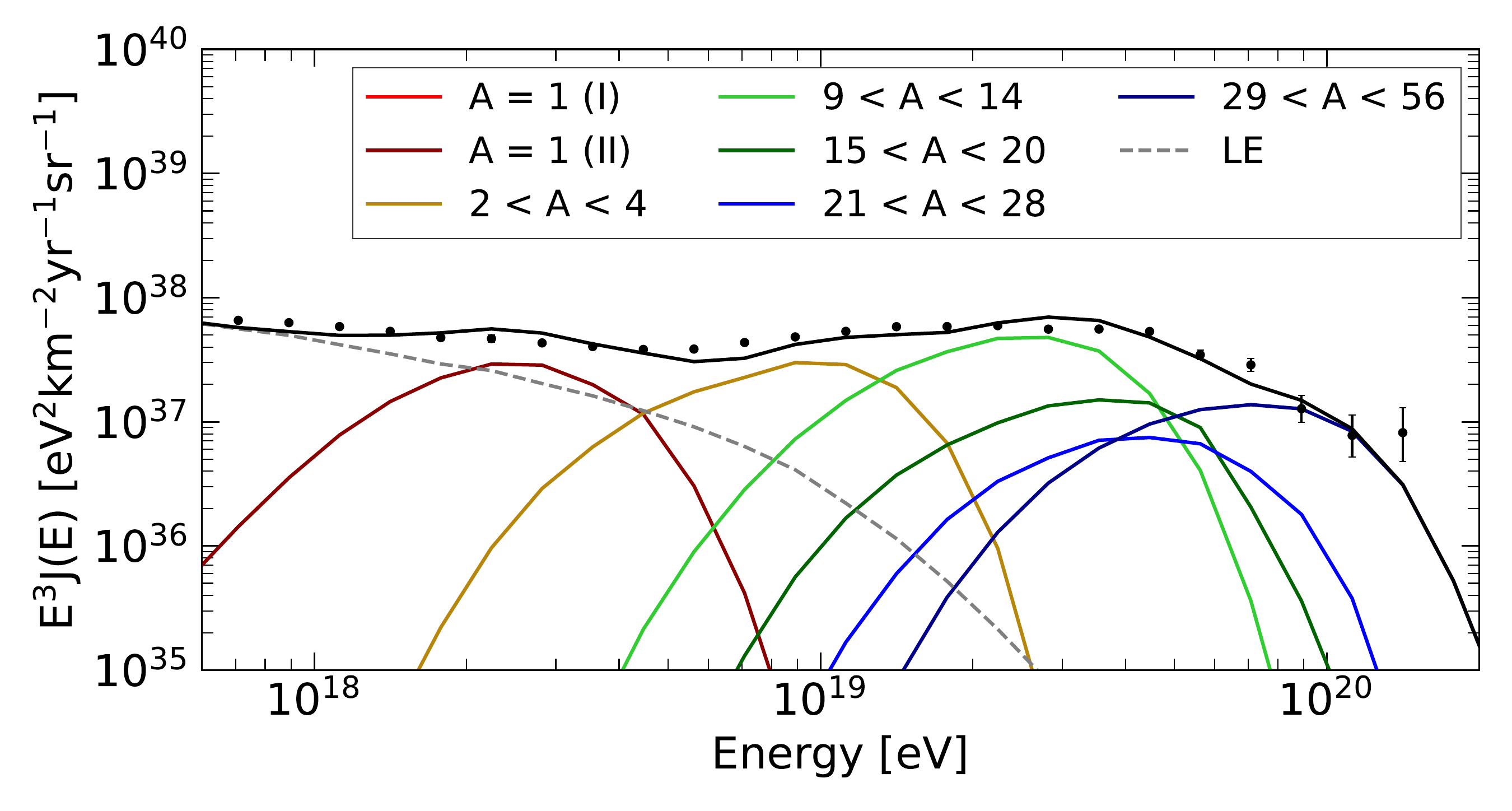}
\includegraphics[width=0.40\linewidth]{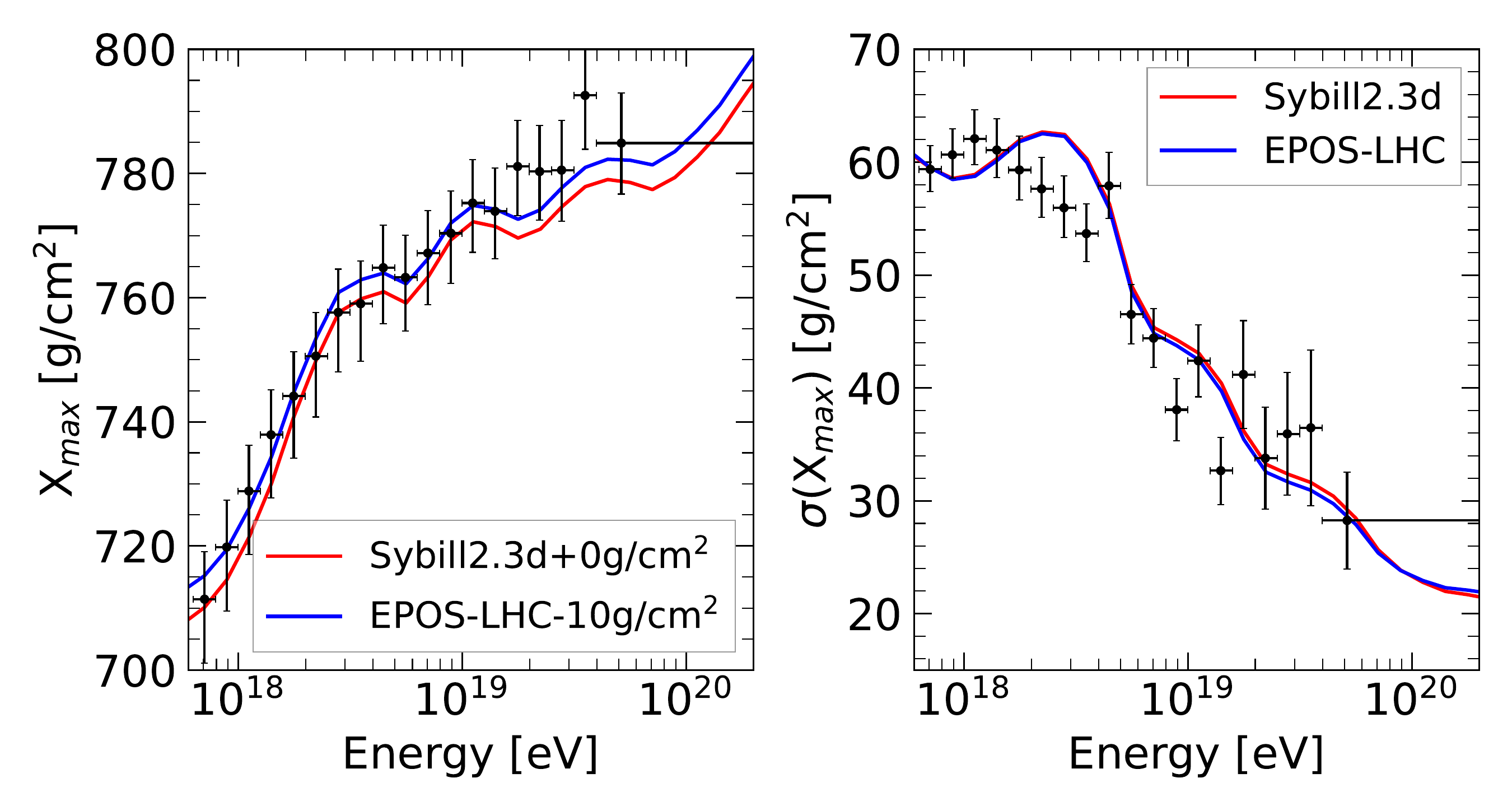}
\caption{UHECR spectrum and corresponding $X_{\rm max}$ moments at Earth, as in Fig.~\ref{fig::spectrum30}, but for $\lambda_B = 50$ Mpc.}        \label{fig::spectrum50}
\end{figure*}

For an even larger suppression scale, $\lambda_B = 50$ Mpc, the production of secondary protons increases further (Fig.~\ref{fig::spectrum50}), while the contribution of heavy nuclei below the instep becomes even more suppressed. The resulting composition is therefore lighter and qualitatively closer to that inferred in the combined fit of~\citet{PierreAuger2023jcapcombinedfitlowenergy} using Sibyll as the hadronic interaction model. In particular, within Sibyll, no shift in the absolute $X_{\rm max}$ scale is required to reproduce the observed $X_{\rm max}$ distribution, whereas EPOS-LHC requires a negative shift.

The LE population required to reproduce the spectrum and composition below the ankle is also broadly consistent with the combined fit to the Auger data \cite{PierreAuger2023jcapcombinedfitlowenergy}: it is dominated by protons and light nuclei.
In particular, we fix the spectral index so as not to overshoot the all-particle spectrum at lower energies ($\lesssim 6 \times 10^{17}$ eV) and therefore obtain the required steepening through the introduction of a maximum-rigidity cutoff, which is not constrained in the combined fit of the Auger data \cite{PierreAuger2023jcapcombinedfitlowenergy}. 

The parameter sets for the benchmark cases are summarized in Table~\ref{tab::Parameters}.

\begin{table}[ht]
\centering
\begin{tabular}{|c|c|c|c|c|c|c|}
\hline
& \multicolumn{2}{c|}{$\lambda_B = 15 \, {\rm Mpc}$} & \multicolumn{2}{c|}{$\lambda_B = 30 \, {\rm Mpc}$} & \multicolumn{2}{c|}{$\lambda_B = 50 \, {\rm Mpc}$} \\
\hline
& HE & LE & HE & LE & HE & LE \\
\hline
$Q_0$ & / & 40 & / & 28 & / & 20 \\
$n_0$  & 2.2 & / & 6.5 & / & 21 & / \\
$R_{\rm max}$ & 6 & 0.3 & 6 & 0.6 & 6 & 1 \\
$\eta_{\rm H}$  & 15\% & 50\% & 0\% & 55\% & 0\% & 50\% \\
$\eta_{\rm He}$ & 15\% & 15\% & 40\% & 15\% & 35\% & 20\% \\
$\eta_{\rm N}$  & 25\% & 20\% & 25\% & 10\% & 40\% & 20\% \\
$\eta_{\rm Ne}$ & 15\% & 5\%  & 20\% & 10\% & 10\% & 10\% \\
$\eta_{\rm Si}$ & 15\% & 5\%  & 5\%  & 10\% & 5\%  & 0\% \\
$\eta_{\rm Fe}$ & 15\% & 5\%  & 10\% & 0\%  & 10\% & 0\% \\
\hline
$\Delta X_{\rm max}^{\rm Syb}$  & \multicolumn{2}{c|}{+20} & \multicolumn{2}{c|}{+10} & \multicolumn{2}{c|}{0} \\
$\Delta X_{\rm max}^{\rm Epos}$ & \multicolumn{2}{c|}{+10} & \multicolumn{2}{c|}{0} & \multicolumn{2}{c|}{-10} \\
\hline
\end{tabular}
\caption{Parameter sets for the benchmark cases discussed in the text. The source number density $n_0$ of the population subject to confinement (HE) is given in units of $10^{-6}$\,Mpc$^{-3}$; the emissivity $Q_0$ of the low-energy population (LE) is given in units of $10^{45}$\,erg\,Mpc$^{-3}$\,yr$^{-1}$; the maximum rigidity is given in EV; and the shift in the absolute $X_{\rm max}$ scale is given in g\,cm$^{-2}$.}
\label{tab::Parameters}
\end{table}

\subsection{Diffuse secondary emission as a consistency check}

\begin{figure*}
\centering
\includegraphics[width=0.40\linewidth]{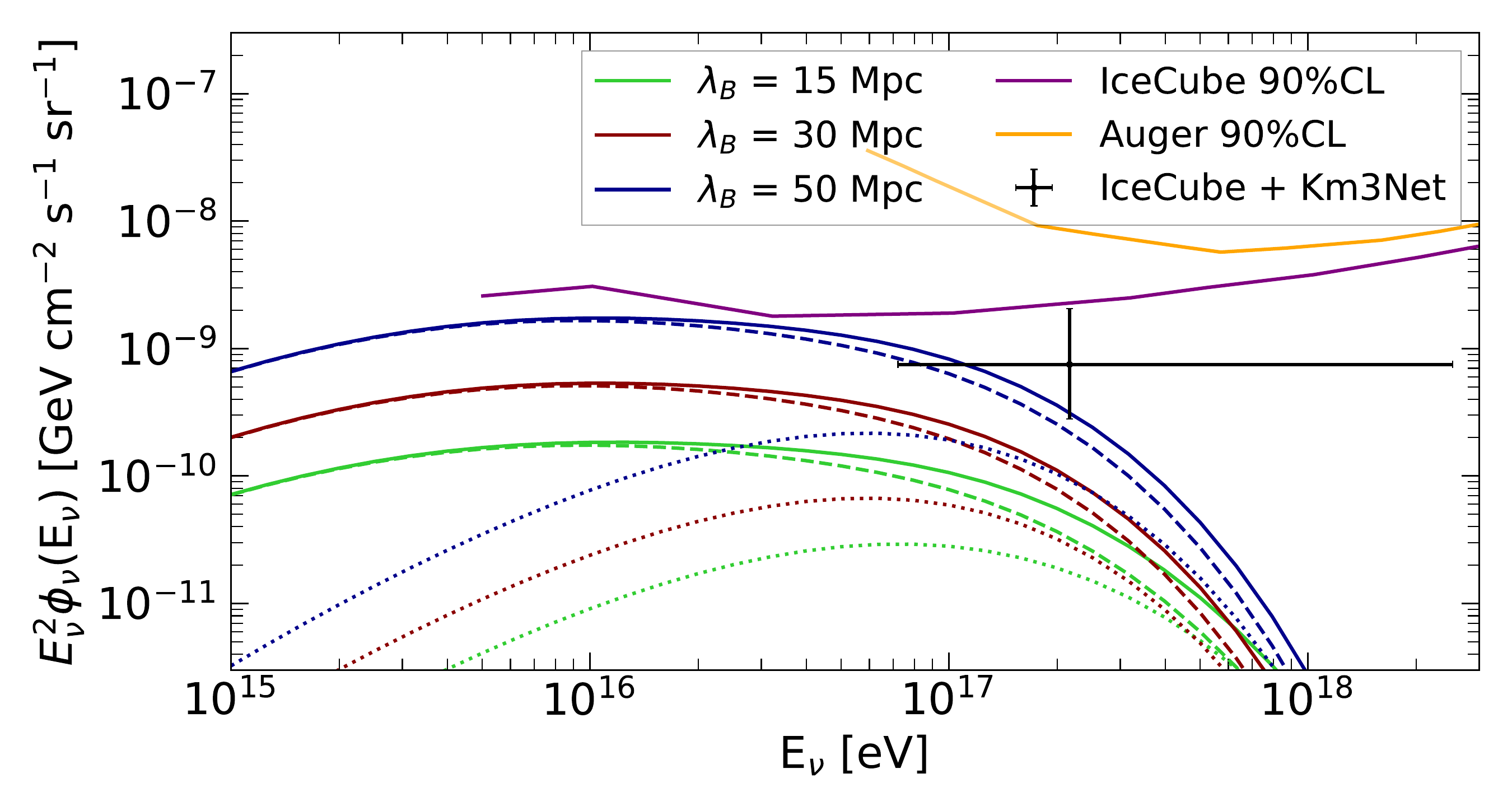}
\includegraphics[width=0.40\linewidth]{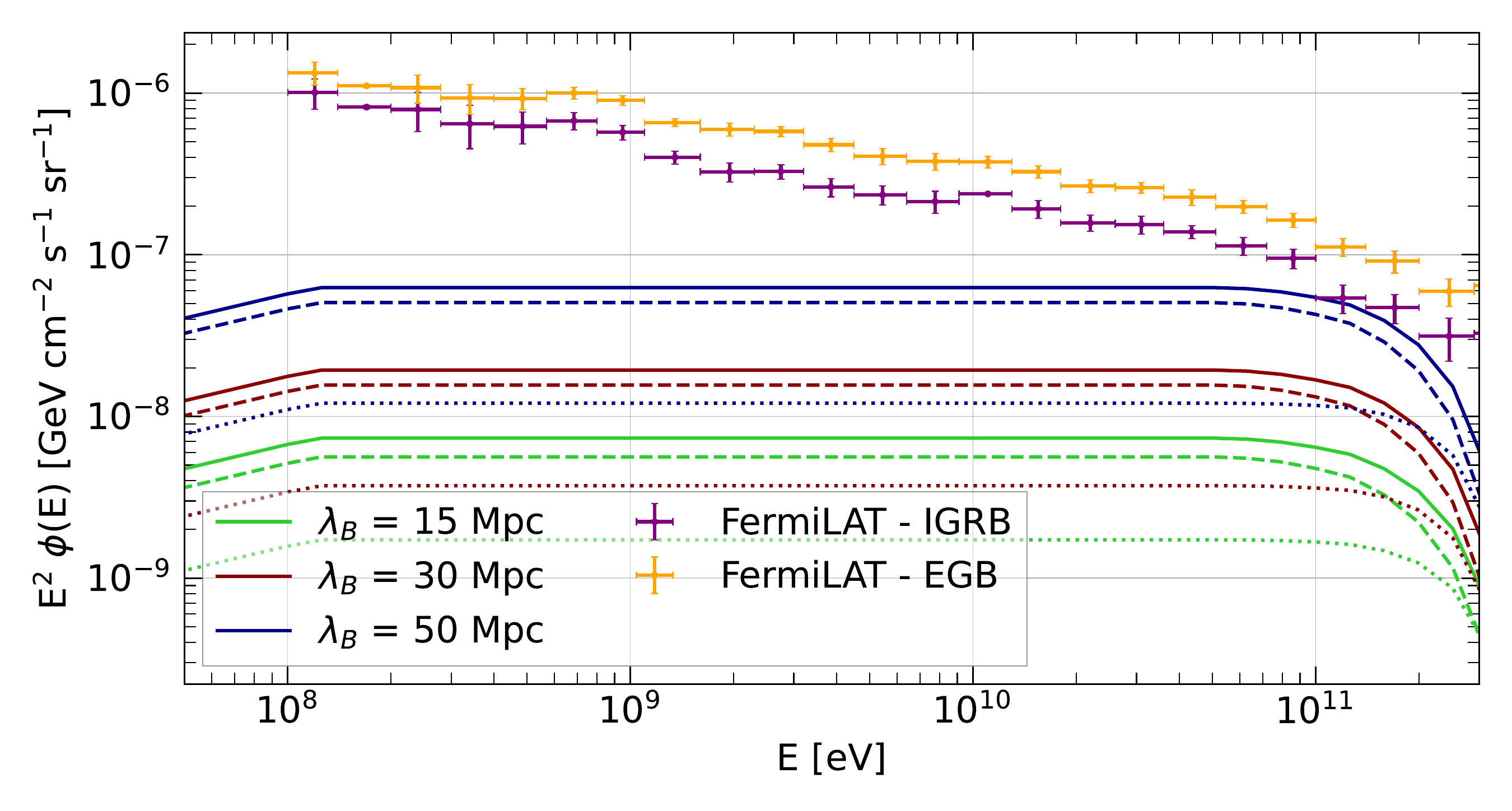}
\caption{\textit{Left}: Diffuse neutrino fluxes associated with the benchmark UHECR spectra shown in Figs.~\ref{fig::spectrum15}, \ref{fig::spectrum30}, and \ref{fig::spectrum50}. The contribution produced inside the confinement region is shown with dashed lines, while the standard cosmogenic contribution from particles after escape is shown with dotted lines. The total flux is shown with solid lines. Model predictions are compared with the latest upper limits from IceCube~\citep{IceCube:2025ezc} and Auger~\citep{PierreAuger2024icrcphotons}, as well as with the flux level inferred from the combined KM3NeT--IceCube analysis~\citep{KM3NET2025prx}. \textit{Right}: Diffuse gamma-ray fluxes for the same benchmark models. The curves represent the cascade emission associated with the UHECR propagation, and are compared with the isotropic gamma-ray background measured by Fermi-LAT~\citep{FermiLat2015apj, FermiLat2016prl}.}
\label{fig::secondaries}
\end{figure*}

The production of secondary cosmogenic neutrinos and gamma rays increases with the confinement timescale, and the corresponding fluxes are shown in Fig.~\ref{fig::secondaries}. As expected, a larger coherence length $\lambda_B$ leads to longer residence times inside the magnetized region, thereby enhancing photodisintegration and photohadronic interactions before escape.

Consistently with previous findings~\cite{Cermenati2026aa}, the benchmark models considered here remain compatible with current neutrino upper limits~\citep{IceCube:2025ezc,PierreAuger2024icrcphotons} for confinement scales up to $\lambda_B \lesssim 50$ Mpc. In all cases, the neutrino flux produced inside the self-confined region dominates over the standard cosmogenic contribution generated during subsequent extragalactic propagation of the released particles. This reflects the fact that, in our scenario, interactions during confinement are more efficient than those occurring after escape over most of the relevant energy range.

In particular, for the case with the largest confinement scale, $\lambda_B = 50$ Mpc, the neutrino flux produced by confined secondary protons reaches a level comparable to that inferred from the KM3NeT event and from recent global-fit estimates~\cite{KM3NET2025prx}. Although this comparison should not be over-interpreted on an event-by-event basis, it is noteworthy that the predicted flux falls in the phenomenologically interesting range suggested by current observations.

Following~\citet{Cermenati2025arxiv}, we also compute the diffuse gamma-ray flux generated through electromagnetic cascading and compare it with Fermi-LAT measurements above $E \gtrsim 50$ GeV. We find that, at the highest energies, the predicted cascade flux approaches the measured isotropic gamma-ray background. This may indicate the onset of tension, especially for the models with the strongest confinement, since a non-negligible fraction of the measured background is expected to originate from unresolved source populations rather than from a truly diffuse cosmogenic component~\citep{Ajello:2015mfa,Lisanti2016apj}. Therefore, while the neutrino bounds still allow confinement scales up to $\lambda_B \sim 50$ Mpc, the diffuse gamma-ray background may already provide a more restrictive test of the most extreme configurations.

\subsection{Astrophysical viability of the self-confinement scenario}

In this section, we examine whether the parameter space identified above can reconcile UHECR observations with more natural acceleration spectra once self-confinement is taken into account.
As we show below, current data leave most of this region viable, while future constraints on the strength and coherence scale of extragalactic magnetic fields, combined with diffuse secondary neutrino and gamma-ray observations, will provide important tests of this scenario.

Our results show that a suppression of the escaping flux at a rigidity \(R \sim 1\)~EV can reproduce the combined fits to the UHECR spectrum and composition~\citep{PierreAuger2017jcapcombinedfit,PierreAuger2023jcapcombinedfitlowenergy} without requiring extremely hard injection spectra.
At the same time, as discussed in~\citep{Unger2015prd}, such a suppression naturally corresponds to a ratio \(\tau_{\rm esc}/\tau_{\rm dis} \sim 10\) for interactions on the average EBL. This implies a substantial production of secondary protons, which helps account for the composition observed below the ankle.

As a benchmark environment for self-confinement, we consider the warm-hot intergalactic medium (WHIM) permeating cosmic filaments, with UHECR sources hosted in galaxy clusters. We adopt a characteristic filament density \(\rho_f \approx 30 \rho_c\) and an ambient magnetic field \(B_f \approx 5\)~nG. The source size is taken to be representative of the central region of galaxy clusters, where most of the galaxies reside, and we assume a fiducial value \(R_s = 300\)~kpc. Within this setup, we determine the combinations of source luminosity and magnetic-field coherence length required to produce a suppression of the escaping flux below the EV scale. The scaling of our results with the environmental properties is given by Eq.~\eqref{eq::rigidity_cutoff}.

Two luminosity-related constraints must be satisfied. First, the source luminosity must exceed
\[
\mathcal{L} \gtrsim L_{\rm min} = \frac{c \Lambda R_s^2 B_0^2}{4},
\]
in order to trigger the non-resonant streaming instability (NRSI)~\citep{Blasi2015prl,Cermenati2026aa}. Physically, this requires the energy density in the UHECR current to exceed that of the pre-existing magnetic field, so that non-resonant perturbations can grow efficiently. Second, the luminosity cannot be too large. If
\[
\mathcal{L} \gtrsim L_{\rm max} = \frac{c \Lambda \pi \rho R_s^2 \lambda_B^2}{T_{\rm age}^2},
\]
the UHECR-driven pressure gradients generate advective outflows that evacuate the flux tube on a timescale shorter than the source age, thereby weakening the confinement effect~\citep{Cermenati2026aa}. This translates into a lower bound on the magnetic-field coherence length, \(\lambda_B > \lambda_{\rm min}(\mathcal{L})\), for a given luminosity.

A further lower bound on \(\lambda_B\) follows from the requirement that the highest-energy particles remain magnetized in the unperturbed field. Their Larmor radius must therefore be smaller than the field coherence length, which implies
\[
\lambda_B \gtrsim \lambda_{\rm magn} = e R_{\rm max} B_0.
\]
On the other hand, the coherence length cannot exceed the typical inter-source distance,
\[
d_s \approx n_0^{-1/3}.
\]

The source density is itself related to the luminosity through the total UHECR emissivity, \(\mathcal{Q} \propto n_0 \mathcal{L}\), which is fixed by the observed flux. To account for this correlation, we adopt as a reference the source density required to fit the data in the \(\lambda_B = 30\)~Mpc case, corresponding to an overall emissivity
\[
\mathcal{Q} \approx 2 \times 10^{46}~{\rm erg~Mpc^{-3}~yr^{-1}}.
\]
Assuming \(n(\mathcal{L}) = \mathcal{Q}/\mathcal{L}\), the inter-source-distance condition provides an upper bound on the coherence length,
\[
\lambda_B \lesssim \lambda_{\rm max}(\mathcal{L}) = n(\mathcal{L})^{-1/3}.
\]

We stress that our reference luminosity is normalized to the power injected into GeV particles. It should therefore be compared with the bolometric luminosity of candidate sources only after accounting for the fraction of the source power converted into UHECRs. The power that actually escapes confinement, in the form of particles above \(\sim 600\)~PeV, is correspondingly smaller. In addition, although part of the energy is redistributed through photodisintegration, another fraction is lost during confinement through Bethe--Heitler interactions of protons. As discussed in the previous section, the remaining interaction channels are subdominant and do not significantly affect our conclusions. Figure~\ref{fig::emissivity} shows that, for our benchmark parameters, the source density required to reproduce the observed spectrum, together with the luminosity fractions associated with the different primary species, implies an emissivity of order \(10^{43}\)–\(10^{44}\)~erg~Mpc\(^{-3}\)~yr\(^{-1}\) for each nuclear species (or mass group) above \(\sim 600\)~PeV. These values are consistent with previous estimates~\citep{PhysRevD.104.043017}.

\begin{figure*}
\centering
\includegraphics[width=0.35\linewidth]{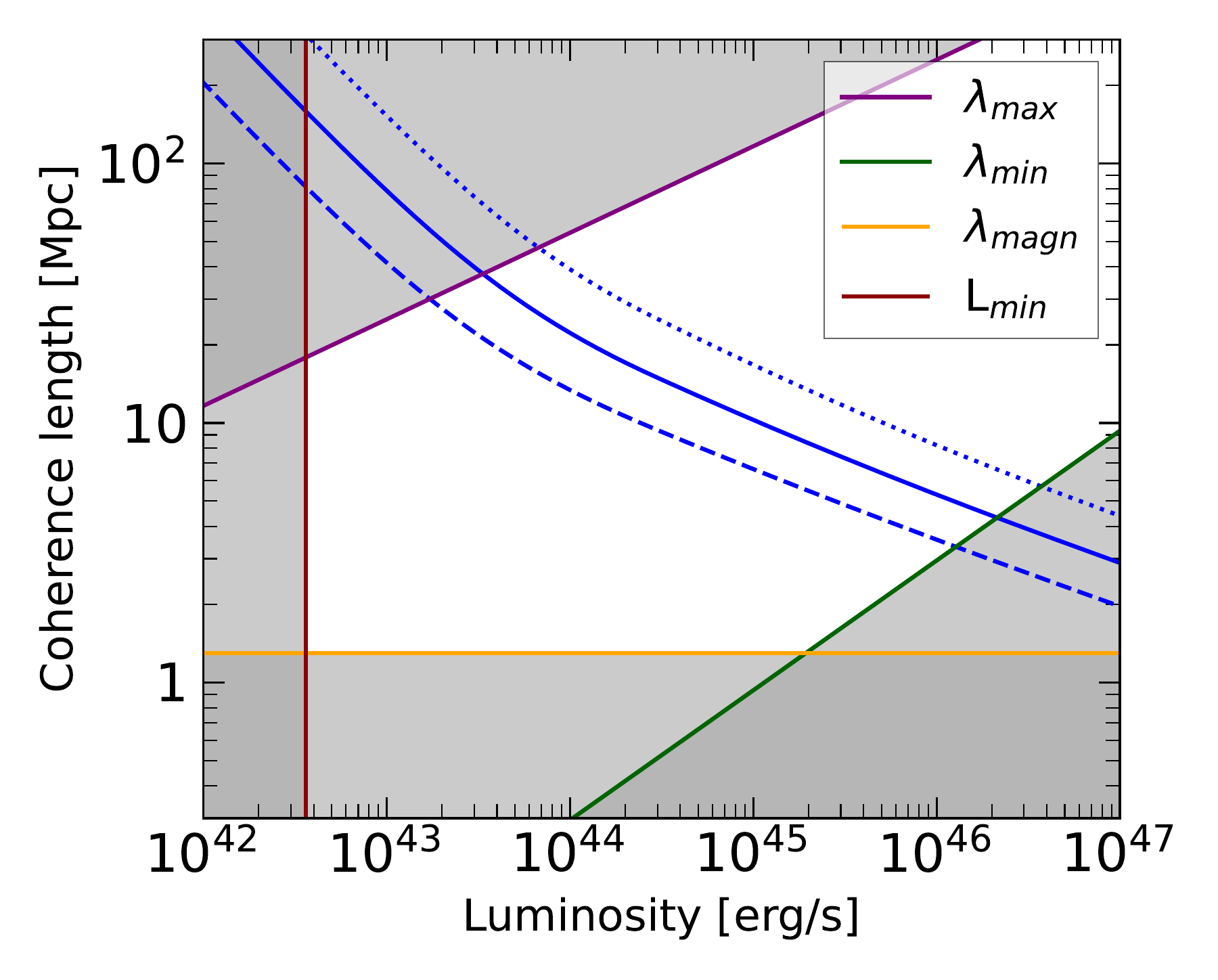}
\includegraphics[width=0.40\linewidth]{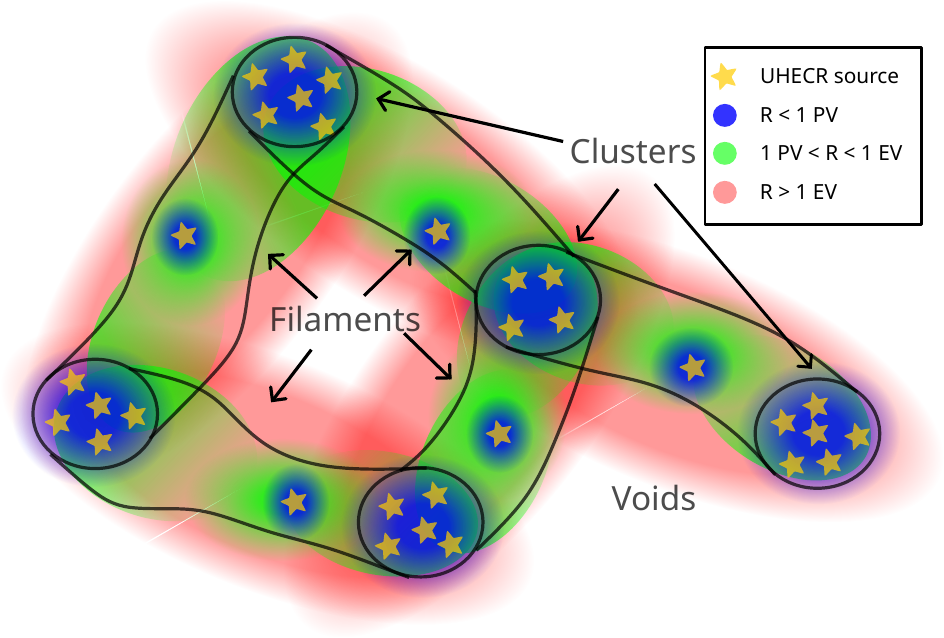}
\caption{{\it Left:} Parameter space in the \(\mathcal{L}\)--\(\lambda_B\) plane relevant for self-confinement in the inter-cluster medium. Shaded regions indicate the excluded parameter space. Blue curves denote the combinations of source luminosity and magnetic-field coherence length required to produce a suppression of the escaping flux at 500~PV (dashed), 1~EV (solid), and 2~EV (dotted). {\it Right:} Schematic representation of self-confinement: UHECR sources are concentrated in galaxy clusters and emit UHECRs that propagate in filaments. Low-rigidity particles (blue) are trapped close to galaxy clusters, while high-rigidity particles (green) produce the NRSI in filaments and only the highest rigidity particles (red) escape from filaments.}
\label{fig::ParSpace}
\end{figure*}

The resulting allowed parameter space is shown in the left panel of Fig.~\ref{fig::ParSpace}. The shaded regions indicate the portions of the \(\mathcal{L}\)–\(\lambda_B\) plane excluded by the conditions discussed above. The blue curves show the coherence length \(\lambda_B(\mathcal{L})\) required, for a given luminosity, to suppress the escaping flux at 500~PV (dashed), 1~EV (solid), and 2~EV (dotted), according to Eq.~\eqref{eq::rigidity_cutoff}. Their different slopes reflect the transport regime of the escaping particles, namely the diffusive, Bohm, and small-angle-scattering regimes~\citep{Cermenati2026aa}.

For the benchmark environment adopted here, a suppression of the UHECR flux at \(R \sim 1\)~EV is achieved for source luminosities in the range
\[
10^{43}~{\rm erg~s^{-1}} \lesssim \mathcal{L} \lesssim 10^{46}~{\rm erg~s^{-1}},
\]
and magnetic-field coherence lengths in the range
\[
5~{\rm Mpc} \lesssim \lambda_B \lesssim 50~{\rm Mpc}.
\]
These values correspond to source number densities of order \(10^{-6}\)–\(10^{-4}\)~Mpc\(^{-3}\) and are consistent with arrival-direction analyses~\citep{Bister2024apj}.

\section{Discussion and conclusions}
\label{discussion}

The Auger measurements of the UHECR spectrum and mass composition are often interpreted as evidence that the population of sources dominating above the ankle must inject an unusually hard spectrum, with a maximum rigidity of a few EV~\citep{PierreAuger2017jcapcombinedfit,PierreAuger2023jcapcombinedfitlowenergy}.

A leading class of explanations attributes this apparent hardening to propagation effects. In particular, it may result from confinement in extended magnetized environments, similarly to what has long been discussed for galaxy clusters~\citep{Berezinsky1997,Ensslin1997apj}, or from the presence of strong extragalactic magnetic fields producing a magnetic-horizon effect~\citep{Aloisio2004apj,Lemoine2005prd,Mollerach2013jcap}. At the same time, composition data suggest that the disintegration of nuclei confined close to, or inside, the source environment may play an important role during propagation. This process modifies the nuclear mixture and naturally produces a peak of secondary protons whose maximum energy scales with the mass of the parent nucleus, rather than with its rigidity, when compared with the spectra of individual nuclear species~\citep{Unger2015prd}.

In this work, we have extended the scenario proposed in previous studies~\citep{Blasi2015prl,Cermenati2026aa}, in which the required confinement is not externally imposed but arises self-consistently from the excitation of the non-resonant streaming instability in the intergalactic magnetic field. When the energy density carried by escaping UHECRs exceeds that of the ambient magnetic field, this instability can generate strong turbulence and substantially modify particle transport. Here we have generalized that framework to the physically relevant case of mixed-composition injection, consistently including the nuclear interaction processes that take place in the confinement region.

Our results show that, under realistic assumptions on UHECR sources and on the surrounding magnetized environment, the turbulence required to delay particle escape up to timescales comparable to or longer than the age of the Universe at rigidities around \(R \sim 1\)~EV can indeed be generated through the growth of the non-resonant streaming instability. At the same time, if one adopts an average EBL level, the additional grammage accumulated during confinement enhances nuclear disintegration and leads to the production of a secondary proton component that can dominate the spectrum below the ankle, thereby helping to reconcile the observed composition with more natural source spectra.

The mechanism appears to be most efficient for sources with a total cosmic-ray power of the order of \(10^{43}\text{--}10^{46}\)~erg~s\(^{-1}\), averaged over the age of the Universe, and for magnetic-field coherence lengths in the range \(\lambda_B \sim 5\text{--}50\)~Mpc. For our reference luminosities, the source density required to match the observed UHECR energy budget is of the order of \(10^{-6}\text{--}10^{-4}\)~Mpc\(^{-3}\), corresponding to mean source separations larger than the field coherence length. Candidate UHECR accelerators such as AGNs~\citep{Fotopoulou2016aa,Lacy2015apj} and starburst galaxies~\citep{Rodighiero2010aa,Condorelli2023prd} may satisfy these requirements if roughly \(10\%\) of their bolometric power is converted into UHECRs. Since such sources are often hosted in galaxy clusters, the cluster environment itself may effectively play the role of the ``source region'' in our model. Formation shocks at the outskirts of galaxy clusters may also satisfy the energetic requirements~\citep{Bohringer2014aa}.

As discussed in~\citep{Cermenati2026aa}, the main limitation of the scenario discussed here is set by the properties of the pre-existing intergalactic magnetic field. Because the maximum rigidity is relatively low, particles remain sufficiently tied to magnetic-field lines over the full energy range of interest. However, for \(R_S \ll 1\)~Mpc, the radial dilution of the escaping current limits the development of the instability on the largest scales, which in turn affects the confinement of particles with rigidities \(R \gtrsim 1\)~EV. This difficulty can be alleviated if the initial field strength is sufficiently large, \(B_0 \gtrsim 1~{\rm nG}\,R_{\rm Mpc}\), so that the instability can grow on the relevant scales within the source lifetime. If such field strengths were ubiquitous throughout the Universe, they could be in tension with Faraday-rotation constraints~\citep{OSullivan2020mnras,Carretti2023mnras,Bondarenko2024arxiv}. However, they remain compatible with current observations if magnetization is concentrated in the filamentary structures of the cosmic web, as suggested by Faraday-rotation and cross-correlation studies involving thermal and synchrotron emission~\citep{Carretti2022mnras,universe10070287}. In this respect, cosmic filaments connecting galaxy clusters appear to be the most natural environment for the mechanism discussed here: they provide higher matter density than voids, and correspondingly weaker observational constraints on the magnetic-field strength.

In the right panel of Fig.~\ref{fig::ParSpace}, we show a schematic representation of the scenario discussed here. UHECR sources are concentrated in galaxy clusters, and UHECRs then propagate in filaments over distances proportional to their rigidity. During such propagation, low-rigidity particles (\( \lesssim \) PV) are subject to small-scale turbulence and remain confined within galaxy clusters for cosmological times. High-rigidity particles (\( \lesssim \) EV) escape into cosmic filaments, where they trigger the NRSI confining themselves. At higher rigidities (\( \gtrsim \) EV), particles propagate diffusively in cosmic filaments, eventually escaping from the self-generated magnetic turbulence and reaching Earth.

Finally, we have shown that particles confined in these magnetized environments produce cosmogenic neutrinos and gamma rays at a level that can exceed the contribution from particles after release into intergalactic space. This opens the possibility of testing the proposed scenario through a multimessenger approach, rather than relying on UHECR spectrum and composition data alone.

Future measurements of the mass composition at the highest energies, in particular with the AugerPrime upgrade of the Pierre Auger Observatory~\citep{Castellina2019epjwc}, together with the improved sensitivity to diffuse UHE neutrinos expected from next-generation observatories such as GRAND, GCOS, and IceCube-Gen2~\citep{GRAND:2018iaj,ahlers2025ideasrequirementsglobalcosmicray,IceCube-Gen2:2020qha}, will be crucial to further probe the viable parameter space of this self-confinement scenario. More generally, they will provide an important opportunity to test whether self-generated turbulence in magnetized large-scale structures plays a fundamental role in shaping the propagation of the highest-energy particles in the Universe.

\begin{acknowledgements}

This work was partially funded by the European Union - NextGenerationEU under the MUR National Innovation Ecosystem grant  ECS00000041 - VITALITY/ASTRA - CUP D13C21000430001. The work of PB was also partially funded by the European Union - Next Generation EU, through PRIN-MUR 2022TJW4EJ.

\end{acknowledgements}

\bibliographystyle{aa}
\bibliography{Cermenati2026}


\begin{appendix}

\section{Advection/Diffusion equation and Leaky Box}
\label{apx::AdvectionDiffusion}

We model particle transport within a flux tube of size $\lambda_B$, under the effects of advection and diffusion. In our picture, particles are emitted at the origin of the flux tube and escape if they reach the opposite boundary, located at a distance $\lambda_B$. The most suitable equation to describe such a situation, neglecting energy losses and the time evolution of the diffusion coefficient, is a one-dimensional advection--diffusion equation of the form:
\begin{equation}
    \frac{\partial n(E,x,t)}{\partial t} + V_A\frac{\partial n(E,x,t)}{\partial x} - D(E)\frac{\partial^2 n(E,x,t)}{\partial x^2} = \frac{Q(E)}{\pi R_s^2}\delta(x).
\label{apx::eq::AdvectionDiffusion}
\end{equation}
The general solution, assuming an initially vanishing density throughout the flux tube, reads:
\begin{equation}
    n(E,x,t) = \int_0^t d\tau \, \frac{Q(E)}{\pi R_s^2 \sqrt{4\pi D(E)(t-\tau)}} {\rm Exp} \left(- \frac{(x - V_A(t-\tau))^2}{4D(E)(t-\tau)} \right),
\label{apx::eq::AD_Density}
\end{equation}
where the exponential term is related to the $\delta$-function in the injection term (Green function). For distances $V_A t \ll x \lesssim \sqrt{4D(E)t}$, the exponential term can be neglected and the integral performed analytically, recovering the leaky box limit:
\begin{equation}
    n(E,x,t) \approx \frac{Q(E)}{\pi R_s^2} \sqrt{\frac{t}{4D(E)}} \approx \frac{Q(E)}{\pi R_s^2 x} \frac{x^2}{4D(E)} \approx \frac{Q(E) \tau_{\rm diff}(E)}{V},
\label{apx::eq::AD_Limit}
\end{equation}
where $V = \pi R_s^2 x$ identifies the volume occupied by particles at time $t$, and $x$ is identified with their diffusive length $l_d$. Therefore, Eq.~\eqref{apx::eq::AD_Limit} is valid at distances $x \approx l_d$. The escaping flux is determined by the particles that reach a distance $\lambda_B$ from the source -- i.e., particles with diffusive lengths $l_d \gtrsim \lambda_B$. In this situation, the exponential term becomes negligible, and the equilibrium density can be approximated with the Leaky Box model. Conversely, in the opposite limit, the instantaneous flux crossing the system boundary, located at a distance $\lambda_B$ from the source, is expected to be suppressed by a factor:
\begin{equation}
    G(E) \approx {\rm Exp} \left(- \frac{(\lambda_B - V_A t)^2}{4D(E)t} \right).
\label{apx::eq::AD:GreenFunction}
\end{equation}
However, to compute the number of interactions experienced by confined particles, the volume they occupy is irrelevant as long as the background target field is uniform. Therefore, the equilibrium density of confined particles can still be determined through the Leaky Box approximation.

\section{UHECR spectrum and composition}
\label{apx::spectrum_and_composition}

In our model, each primary UHECR injected by the source undergoes disintegration during the confinement process; therefore, we expect all nuclei up to iron to escape the confinement region and reach Earth. \\
To calculate the resulting spectra of each nucleus at Earth, we run a dedicated simulation with \textit{Simprop} for each nucleus included in our analysis. We consider six different primary species injected by the source: protons, Helium (He), Nitrogen (N), Neon (Ne), Silicon (Si), and Iron (Fe). Hence, the total output power in UHECRs from each source is divided among the six representative elements. The resulting escaping flux of UHECRs originating from the disintegration of a given primary must be normalized by the corresponding fraction of the source luminosity $\eta_i$, with the constraint:
\begin{equation}
    \sum_i \eta_i = 1.
\label{apx::eq::power_fractions}
\end{equation}
For each primary nucleus with mass $A_{\rm src}$, the computation of the emissivity of each escaping nucleus type, with mass $A_{\rm IGM} \leq A_{\rm src}$, is described in Section~\ref{sec:confinementandinteractions}. We define the emissivity of nuclei with mass $A_{\rm IGM}$, originating in the disintegration chain of primary nuclei with mass $A_{\rm src}$, as:
\begin{equation}
    \widetilde{Q}_{A_{\rm src}}(E,z, A_{\rm IGM}) = \widetilde{Q}_{\rm I/II}^{(A_{\rm IGM},Z_{\rm IGM})}(E,z).
\label{apx::eq::SimPropEmissivity}
\end{equation}
The spectrum at Earth of nuclei with mass $A_{\rm Earth}$, originating from primary nuclei with mass $A_{\rm src}$, is given by:
\begin{equation}
    E_{\rm obs}^3 J_{A_{\rm src}}(E_{\rm obs}, A_{\rm Earth}) = \frac{c}{4\pi} \times \sum_{A_{\rm IGM} = A_{\rm Earth}}^{A_{\rm IGM} = A_{\rm src}} \frac{N_{\rm Earth}(E_{\rm obs}, A_{\rm Earth})}{N_{\rm tot}} W(E_{\rm obs}, E_{\rm inj}, z_{\rm inj}) Q_{A_{\rm src}}(E_{\rm inj}, z_{\rm inj}, A_{\rm IGM}),
\label{apx::eq::SimPropFlux}
\end{equation}
where $N_{\rm tot} = 100000$ is the total number of events generated with injected mass $A_{\rm IGM}$, and $N_{\rm Earth}(E_{\rm obs}, A_{\rm Earth})$ is the total number of particles reaching Earth with energy $E_{\rm obs}$ and mass $A_{\rm Earth}$. The reweighting function $W$ accounts for the energy binning, the expansion of the Universe, and the injection spectral slope, following the approach adopted in~\citep{Unger2015prd}:
\begin{equation}
    W(E_{\rm inj}, E_{\rm obs}, z_{\rm inj}) = {\rm ln}\left( \frac{10^{21}}{10^{17}} \right) E_{\rm obs}^3 \left| \frac{dt}{dz} \right| \frac{E_{\rm inj}}{E_{\rm obs} {\rm ln}(10) \Delta{\rm Log}(E)}.
\label{apx::eq::SimPropWeight}
\end{equation}
The overall spectrum of particles with mass $A_{\rm Earth}$ is then given by the sum of all spectra generated by each considered primary, normalized to the respective luminosity fraction:
\begin{equation}
    E^3 J_{\rm Tot}(E, A_{\rm Earth}) = \sum_{A_{\rm src}} \eta_{A_{\rm src}} E^3 J_{A_{\rm src}}(E, A_{\rm Earth}).
\label{apx::eq::SimPropIndividualSpectra}
\end{equation}
The all-particle spectrum is then given by the sum over the spectra of individual species:
\begin{equation}
    E^3 J_{\rm Tot}(E) = \sum_{A_{\rm Earth}} E^3 J_{\rm Tot}(E, A_{\rm Earth}).
\label{apx::eq::SimPropSpectrum}
\end{equation}
To interpret the air-shower maximum depth in the atmosphere, and its variance, produced by our model spectra, we consider the parameterization described in~\citet{Evoli2026aph}. The moments of the shower-maximum distribution are related to the average ${\rm ln}(A)$ and ${\rm ln}(A)^2$, which can be computed as:
\begin{equation}
    \begin{split}
        &<{\rm ln}(A)>(E) = \frac{\sum_{A_{\rm Earth}} \left({\rm ln}(A_{\rm Earth}) J^{\rm Tot}(E, A_{\rm Earth})  \right) }{J^{\rm Tot}(E)}, \\
        &<{\rm ln}(A)^2>(E) = \frac{\sum_{A_{\rm Earth}} \left(({\rm ln}(A_{\rm Earth}))^2 J^{\rm Tot}(E, A_{\rm Earth})  \right) }{J^{\rm Tot}(E)}.
    \end{split}
\label{apx::eq::AveragedMass}
\end{equation}

\section{Secondary particles from confined protons}
\label{apx::secondaries_confined}

We account for the propagation of secondary cosmogenic neutrinos and gamma-rays analytically. Neutrinos undergo only adiabatic energy losses, and their spectrum at Earth can be easily calculated as~\citep{Cermenati2025arxiv}:
\begin{equation}
    J_\nu(E_\nu) = \frac{c}{4 \pi} \int_0^{z_{\rm max}} dz_g \, \left| \frac{dt}{dz_g} \right| \, \widetilde{Q}_\nu(E_\nu (1+z_g), z_g) (1 + z_g),
\label{apx::eq::neutrino_spectrum}
\end{equation}
where $\widetilde{Q}_\nu(E_\nu, z)$ is the comoving emissivity of cosmogenic neutrinos from confined particles, and the factor $(1+z_g)$ accounts for adiabatic energy losses during propagation. Leptons and gamma-rays, conversely, undergo strong interactions (inverse Compton scattering and pair production) with a typical timescale much shorter than any propagation timescale across the Universe. The universality approximation made in the calculation of the emissivity (Eq.~\eqref{eq::GammaRay_Emissivity}) treats the full development of the cascade as instantaneous~\citep{Berezinsky2016prd}; therefore, the remnant gamma-rays undergo only adiabatic energy losses, as cosmogenic neutrinos do, and their spectrum at Earth can be easily computed as~\citep{Cermenati2025arxiv}:
\begin{equation}
    J_\gamma(E_\gamma) = \frac{c}{4 \pi} \int_0^{z_{\rm max}} dz_g \, \left| \frac{dt}{dz_g} \right| \, \widetilde{Q}_\gamma(E_\gamma (1+z_g), z_g) (1 + z_g),
\label{apx::eq::gammaray_spectrum}
\end{equation}

\section{Secondary particles from released protons}
\label{apx::secondaries_released}

The spectrum of cosmogenic neutrinos and gamma-rays produced by released particles can be calculated with the same methodology described in Appendix~\ref{apx::secondaries_confined}, but replacing the emissivity with that of released protons, which in turn depends on their equilibrium density $\widetilde{n}_p(E,z)$. This quantity can be computed analytically following the methodology presented in~\citep{Berezinsky2006prd,Cermenati2025arxiv}. The emissivity of cosmogenic neutrinos is given by:
\begin{equation}
    \widetilde{Q}_\nu (E_\nu,z) = \sum_i \left( \int_{E_\nu}^{+\infty} \frac{dE_p}{E_p} \widetilde{n}_p(E_p, z) \times R_\pi^i(E_p, E_\nu, z) \right),
\label{apx::eq::CosmoNeutrino_Emissivity}
\end{equation}
where the differential production rate of neutrinos, through photopion production, is defined as:
\begin{equation}
    R_\pi(E_p, E_\nu, z) = \int_{\epsilon_{\rm th}(E_p)}^{+\infty} d\epsilon \, n_b(\epsilon, z) \, \Phi_\pi(E_p, E_\nu, \epsilon).
\label{apx::eq::PhotopionRate}
\end{equation}
Here, $\Phi$ is the parametric function described in~\citep{Kelner2008prd}. The overall emissivity is the sum over the neutrino flavours $i = \nu_\mu, \bar{\nu}_\mu, \nu_e, \bar{\nu}_e$. Analogously, the total energy emitted per unit time into EM cascades can be calculated by integrating over the emitted spectrum of secondary electrons, positrons, and gamma-rays through photopion production:
\begin{equation}
    \widetilde{\Omega}_\pi(z) = \sum_{i = e^-, e^+, \gamma} \int_0^{+\infty} dE_i \, E_i \, \int_{E_i}^{+\infty} \frac{dE_p}{E_p} \, \widetilde{n}_p(E_p, z) \times R_\pi^i(E_p, E_i, z).
\label{apx::eq::CosmoGamma_PionEnergyDensity}
\end{equation}
Conversely, pair production only produces pairs that will undergo the cascading process; therefore, all the energy lost by protons through pair production would feed electromagnetic cascades:
\begin{equation}
    \widetilde{\Omega}_{ee}(z) = \int dE_p \, \widetilde{n}_p(E_p, z) b(E_p, z),
\label{apx::eq::CosmoGamma_PairEnergyDensity}
\end{equation}
where $b(E_p, z) = -dE/dt$ is calculated following~\citep{Chodorowski1992apj,Berezinsky2006prd} for pair production only. The emissivity of gamma-rays can be computed as in Eq.~\eqref{eq::GammaRay_Emissivity}, accounting for the emitted spectrum of remnant gamma-rays through the universality approximation. The spectrum of cosmogenic neutrinos and gamma-rays at Earth is then obtained straightforwardly as in Eqs.~\eqref{apx::eq::neutrino_spectrum} and~\eqref{apx::eq::gammaray_spectrum}.
    
\end{appendix}

\end{document}